\documentclass[showpacs,preprintnumbers,amsmath,amssymb,floatfix]{revtex4}
\usepackage[german,english]{babel}
\usepackage{graphicx}
\usepackage{amsmath}
\usepackage{amsfonts}
\usepackage{amssymb}
\usepackage{epsfig}
\usepackage[hang,nooneline]{subfigure}
\newcommand{\insertplot}[5]{\begin{figure}
 \hfill\hbox to 0.05in{\vbox to #5in{\vfill
 \inputplot{#1}{#4}{#5}}\hfill}
 \hfill\vspace{-.1in}
 \caption{#2}\label{#3}
 \end{figure}}
\newcommand{\inputplot}[3]{
 \special{ps: plotfile #1}
}
 
\newcounter{fig}

\begin{document}

\title{\bf Wormholes Threaded by Chiral Fields}
\author{{\bf Efstathios Charalampidis}}
\email[{\it Email:}]{echarala@auth.gr}
\author{{\bf Theodora Ioannidou}}
\email[{\it Email:}]{ti3@auth.gr}
\affiliation{
Department of Mathematics, Physics and Computational Sciences, Faculty of Engineering,\\
Aristotle University of Thessaloniki 
Thessaloniki 54124,  Greece
}
 
\author{{\bf  Burkhard Kleihaus}}
\email[{\it Email:}]{b.kleihaus@uni-oldenburg.de}
\author{{\bf and Jutta Kunz}}
\email[{\it Email:}]{jutta.kunz@uni-oldenburg.de}
\affiliation{
Institut f\"ur Physik, Universit\"at Oldenburg, Postfach 2503\\
D-26111 Oldenburg, Germany
}

\date{\today}
\pacs{04.20.JB, 04.40.-b}

\begin{abstract}
We consider Lorentzian wormholes with a phantom field and chiral matter fields.
The chiral fields are described by the non-linear sigma model {\it with or without}
a Skyrme term.
When the gravitational coupling of the chiral fields is increased,
the wormhole geometry changes. 
The single throat is replaced by a double throat
with a belly inbetween. 
For a maximal value of the coupling, the radii of
both throats reach zero. Then the interior part pinches off,
leaving a closed universe and two (asymptotically) flat spaces.
A stability analysis shows
that all wormholes threaded by chiral fields 
inherit the instability of the Ellis wormhole. 
\end{abstract}

\maketitle

\section{Introduction}

Einstein's theory of general relativity describes
gravitational phenomena ranging from planetary motion to cosmology.
Among the plethora of solutions of the Einstein
equations there are also the wormhole solutions \cite{Visser:1995cc}.
Wormholes are a hypothetical topological feature of space-time that correspond to 
 ``tunnels" through space-time. 
In 1935 \cite{Einstein:1935tc} the ``Einstein-Rosen bridge''
was discovered as a feature of the Schwarzschild geometry.
It represents a space-time manifold consisting of two asymptotically flat
universes connected by a throat. Wormholes that connect two distant regions
of our own universe were discussed by Wheeler in the
1950's \cite{Wheeler:1957mu,Wheeler:1962}.  However, it was shown that
this kind of wormholes would collapse if one tried to pass them 
\cite{Kruskal:1959vx,Fuller:1962zza,Redmount:1985,Eardley:1974zz,Wald:1980nk}.

Wormholes which could actually be crossed in both directions, are known as {\it traversable wormholes}.
Traversable wormholes were considered in \cite{Morris:1988cz}
when a new kind of matter was coupled to gravity, whose
energy-momentum tensor would violate all (null, weak and strong) energy
conditions. A suitable candidate would be a phantom field, i.e. a scalar field with
a reversed sign in front of its kinetic term
\cite{Ellis:1973yv,Ellis:1979bh,Bronnikov:1973fh,Kodama:1978dw,ArmendarizPicon:2002km}.
Wormholes with such phantom fields have been studied in a variety of settings,
including stars and neutron stars 
\cite{Dzhunushaliev:2011xx,Dzhunushaliev:2012ke,newpap}.
However, the presence of a phantom field is {\it not} necessary to obtain wormholes.
For example, they occur in  theories  when instead of Einstein gravity higher curvature terms are considered, such as $f(R)$-theories or
Einstein-Gau\ss -Bonnet-dilaton theories
\cite{Hochberg:1990is,Fukutaka:1989zb,Ghoroku:1992tz,Furey:2004rq,Bronnikov:2009az, Kanti:2011jz,Kanti:2011yv}.

In this paper, we investigate wormhole solutions when  a phantom field
and  chiral fields are minimally coupled to Einstein gravity.
For the chiral fields we first take the non-linear sigma model, and 
subsequently allow for the presence of a 
higher order term, i.e. a Skyrme term.
Our motive is based on the previous observations  that {\it the presence of 
non-Abelian fields can lead to new interesting
gravitational phenomena, since
``hairy black holes'' were discovered in the 
Einstein-Skyrme model} \cite{Luckock:1986tr,Droz:1991cx}.

Here we restrict to static spherically symmetric wormhole solutions
connecting two asymptotically flat universes.
The solutions are characterized by three parameters: the coupling 
constant to
gravity $\alpha$,  the radius of the throat of the wormhole, and the value of the 
chiral fields at the throat. 
In particular, we study the dependence of the wormhole solutions on these
parameters. Thereby we find 
that the geometry of the wormhole changes as the gravitational
coupling is increased. 
Instead of possessing a single minimal radius, the wormhole develops
a maximal radius surrounded by two minimal radii, i.e., a double throat forms.
At a critical coupling, the two throats reach zero size.
Then the space-time inbetween the two throats
forms a closed universe, which
pinches off from the two exterior universes.

Our main focus is on 
wormholes that are symmetric under the interchange of the two
asymptotically flat universes. We consider these solutions
as chiral configurations localized in the vicinity of the throat.
In the case of a Skyrmion, they might give us an idea of the mechanism taken place
when such an extended particle is passing the throat.
Besides the symmetric wormholes we also consider
non-symmetric wormholes, by changing the boundary
condition of the chiral fields at the throat.

Stability is an essential question for traversable wormholes. 
Recently, it was shown
that wormholes minimally coupled to a phantom field possess an
unstable mode \cite{Gonzalez:2008wd,Bronnikov:2011if}.  
However, self-gravitating Skyrmions possess a conserved topological charge 
and are stable  \cite{Heusler:1991xx};
and likewise, black holes with Skyrmionic hair
are linearly stable \cite{Heusler:1992av}.
Thus one may ask, whether this stability exists also for wormholes permeated by chiral fields. 
After all they {\it do} possess a conserved topological charge.
However, they may also inherit the unstable mode of the 
Ellis wormhole \cite{Ellis:1973yv}.  
To answer this question,
we study small spherically symmetric perturbations of the 
wormholes with chiral fields. 
We impose a harmonic time-dependence on the perturbations 
and find that all wormholes retain the unstable mode
of the Ellis wormhole.
Thus they are all, linearly unstable.

The outline of our paper is as follows: In section II, we present
the action, the Ans\"atze and the field equations.
We discuss the main wormhole features in section III.
The numerical results for the symmetric wormholes 
are presented in section IV. 
Section V is devoted to the stability analysis of these solutions.
We report the main results for the non-symmetric wormholes
in section VI, and give the conclusions in section VII.

\section{Action and Field Equations}

\subsection{Action}

We consider Einstein gravity coupled to a phantom field and ordinary matter fields. 
The action 
\begin{equation}
S=\int \left[ \frac{1}{16\pi G}{\cal R} + 
{\cal L}_{\rm ph} +{\cal  L}_{\rm ch} + {\cal L}_{\rm sk} \right] \sqrt{-g}\  d^4x  
 \label{action}
\end{equation}
consists of the Einstein-Hilbert action
with curvature scalar $\cal R$, Newton's constant $G$ and determinant of the 
metric $g$, and of the corresponding matter contributions. 
These are 
the Lagrangian of the phantom field $\phi$,
\begin{equation}
 {\cal L}_{\rm ph} = \frac{1}{2}\partial_\mu \phi\partial^\mu \phi \ ,
\label{lphi}
\end{equation}
and  the non-linear sigma model Lagrangian
\begin{equation}
{\cal L}_{\rm ch} = \frac{\kappa^2}{4}{\rm Tr}\left\{L_\mu L^\mu \right\}
\label{lch}
\end{equation} 
where $L_\mu = \partial_\mu U U^\dagger$ and $\kappa$ is a coupling constant.
The chiral matrix $U$  is a function on the space-time manifold taking 
values in the Lie group SU(2).
The last term in the action represents the Skyrme term 
\begin{equation}
{\cal  L}_{\rm sk} =\frac{1}{32 e^2}{\rm Tr}\left\{F_{\mu\nu} F^ {\mu\nu}\right\} \ 
\label{lsky}
\end{equation} 
with $F_{\mu\nu} = \left[L_\mu,L_\nu\right]$ and $e$ a coupling constant.

Variation of the action with respect to the metric
leads to the Einstein equations
\begin{equation}
G_{\mu\nu}= {\cal R}_{\mu\nu}-\frac{1}{2}g_{\mu\nu}{\cal R} = 8\pi G T_{\mu\nu}
\label{ee} 
\end{equation}
with stress-energy tensor
\begin{equation}
T_{\mu\nu} = g_{\mu\nu}{{\cal L}}_M
-2 \frac{\partial {{\cal L}}_M}{\partial g^{\mu\nu}} \ ,
\label{tmunu} 
\end{equation}
where ${\cal L}_{\rm M} = 
{\cal L}_{\rm ph}+{\cal L}_{\rm ch} +{\cal L}_{\rm sk} $ is the matter Lagrangian.

\subsection{Ans\"atze}

For the study of static spherically symmetric wormhole solutions
an appropriate choice for the line element would be
\begin{equation}
ds^2 = -A^2 dt^2 +d\eta^2 + R^2d\Omega^2 \ ,
\label{lineel}
\end{equation}
where $d\Omega^2=d\theta^2 +\sin^2\theta d\varphi^2$ 
denotes the metric of the unit sphere, while $A$ and $R$ are functions of 
$\eta$.
Note that the coordinate $\eta$ takes positive and negative 
values, i.e. $-\infty< \eta < \infty$. The limits $\eta\to \pm\infty$
correspond to two disjoint asymptotically flat regions.

We parametrize the chiral matrix as 
\begin{equation}
U = \cos F + i\sin F \ \vec{e}\cdot \vec{\tau} \ ,
\label{skyrmU}
\end{equation}
with the unit vector field $\vec{e}$
\begin{equation}
\vec{e} = 
\left(\sin\theta \cos\varphi , \sin\theta \sin\varphi, \cos\theta \right) \ ,
\label{vece}
\end{equation}
and the vector of Pauli matrices $\vec{\tau}$.
The chiral profile function $F$ is a function of $\eta$.

\subsection{Einstein and Matter Field Equations}

Substitution of the above Ans\"atze into the Einstein equations 
$G_\mu^\nu=8\pi G T_\mu^\nu$ yields
\begin{eqnarray}
\lefteqn{\frac{2 R R'' +R'^2 -1 }{R^2} = }
\nonumber\\
 & & 4 \pi G \left\{\phi'^2 - \frac{\kappa^2(R^2 F'^2 + 2\sin^2 F)}{R^2}
                        -\frac{\sin^2 F (2R^2 F'^2 + \sin^2 F)}{e^2 R^4}
	\right\}
\label{eeq_tt}\\
\lefteqn{\frac{A R'^2 -A + 2R R' A'}{AR^2} = }\nonumber\\
 & & 
4 \pi G \left\{-\phi'^2 + \frac{\kappa^2(R^2 F'^2 - 2\sin^2 F)}{R^2}
                        +\frac{\sin^2 F (2 R^2 F'^2 - \sin^2 F)}{e^2 R^4}
	\right\}
\label{eeq_rr}\\
\lefteqn{\frac{A R'' + A' R' +A'' R}{AR} = }\nonumber\\
 & & 
4 \pi G \left\{\phi'^2 - \kappa^2 F'^2  
                        +\frac{\sin^4 F}{e^2 R^4 }
	\right\}
\label{eeq_oo}			
\end{eqnarray}
for the $tt$, $\eta\eta$ and $\theta\theta$ components, respectively.
(The $\varphi\varphi$ and $\theta\theta$ equations are equivalent.) 

The equations for the chiral profile function and the phantom field are obtained 
from the variation of the action with respect to $F$ and $\phi$, respectively.
They read 
\begin{eqnarray}
\left[\left(\kappa^2 R^2+\frac{2\sin^2 F}{e^2}\right)A F'\right]'
& = & \frac{A \sin 2F}{R^2}
\left(\kappa^2 R^2 +\frac{R^2 F'^2+\sin^2 F}{e^2}\right)\ ,
\label{eqSk}\\
\left[ A R^2 \phi'\right]' &=& 0\ .
\label{eqPh}
\end{eqnarray}
The last equation yields 
\begin{equation}
\phi' = \frac{D}{A R^2} \ ,
\label{phip}
\end{equation}
where the constant $D$ is related to the scalar charge. 
Thus, the phantom field can be eliminated from the Einstein equations
by the substitution $\phi'^2 = D^2/A^2 R^4$.

Next we introduce dimensionless quantities via scaling
\begin{equation}
\eta = \frac{\tilde{\eta}}{\kappa} \ , \ \ \ 
R=\frac{\tilde{R}}{\kappa} \ , \ \ \ 
\phi = \kappa \tilde{\phi} \ , \ \ \ 
4 \pi G = \frac{\alpha}{\kappa^2}\  . 
\label{scale}
\end{equation}
This is equivalent to setting $\kappa=1$  and $4 \pi G =\alpha$.
Finally, we rename $\tilde{\eta}=\eta$, $\tilde{R}=R$, $\tilde{\phi}=\phi$
to simplify the notation. 

We observe that adding Eq.~(\ref{eeq_rr}) to Eq.~(\ref{eeq_tt}) 
and Eq.~(\ref{eeq_oo}) eliminates the $\phi'^2$ term. The resulting 
equations can be cast in the form
\begin{eqnarray}
R'' & = & \frac{A - A R'^2 -R R'A'}{AR} 
- \alpha\left[\frac{2\sin^2 F}{R}+\frac{\sin^4 F}{e^2 R^3}\right] \ ,
\label{eqR}\\
A'' & = & -\frac{2 R'A'}{R} 
+ \alpha\sin^2 F A\frac{2 R^2 F'^2+ \sin^2 F}{e^2 R^4}  \ ,
\label{eqA}
\end{eqnarray}
which form together with Eq.~(\ref{eqSk}) a system of second order
ODEs to be solved numerically.
The scalar charge $D$ for the solutions can be obtained from Eq.~(\ref{eeq_rr}),
\begin{equation}
 D^2 = 
-\frac{1}{\alpha}AR^2\left(A R'^2 -A + 2R R' A'\right)
+A^2\left\{ \left(R^2 F'^2 - 2\sin^2 F\right)R^2
                        +\frac{\sin^2 F}{e^2} \left(2 R^2 F'^2 - \sin^2 F\right)
	\right\} \ .
\label{eqd}
\end{equation}
The condition $D=const$ is used to monitor the quality of the
numerical solutions.

\section{Wormhole Geometry}

\subsection{Geometry of  the Throat}

To obtain wormhole solutions,
we assume that the function $R$ does not possess any zero.
For asymptotically flat solutions, $R$ behaves like $|\eta|$ in the 
asymptotic regions. Consequently, $R$ possesses (at least) one minimum. 
Suppose $R'(\eta_0)=0$  for some $\eta_0$ and $R(\eta_0)=r_0$.
Then it follows from Eq.~(\ref{eqR}) that
\begin{equation}
R''(\eta_0) = \frac{1}{r_0} 
\left(1- \alpha\sin^2 F_0\left[2 + \frac{\sin^2 F_0}{e^2 r_0^2}\right]\right) 
=\frac{1}{r_0} \left(1- \alpha/\alpha_{\rm cr}\right) 
\ ,
\end{equation}
where $F_0 = F(\eta_0)$, and we defined
\begin{equation}
\alpha_{\rm cr}= \frac{1}{\sin^2 F_0\left[2 + \frac{\sin^2 F_0}{e^2 r_0^2}\right]}\ .
\end{equation}
Thus, $R$ possesses a minimum at $\eta_0$ if $\alpha < \alpha_{\rm cr}$
and  a maximum if  $\alpha > \alpha_{\rm cr}$.
The first case $\alpha < \alpha_{\rm cr}$ corresponds to the typical
wormhole scenario: a surface of minimal area separates two 
asymptotically flat regions. 
In the second case the maximum is a local maximum since
in the asymptotic regions $R= |\eta|$. This implies that there are 
(at least) two minima of $R$, one for $\eta < \eta_0$ and another one for $\eta > \eta_0$.
The wormhole then has a double throat.
For a sequence of alternating minima and maxima
the spatial hyper-surfaces would also possess two asymptotically flat regions, but they would be
separated by a sequence of throats, thus a true multi-throat wormhole would arise.

\subsection{Boundary Conditions}

For the position of the throat - or beyond $\alpha_{\rm cr}$ of the maximal area surface,
i.e., the equator - 
we can choose without loss of generality
$\eta_0=0$. 
At this extremal surface - the throat or the equator -
we then impose the conditions
\begin{equation}
R(0)= r_0 \ , \ \ \ 
R'(0)=0 \ ,  \ \ \ 
F(0)=F_0 \ . 
\label{bcthro}
\end{equation}
The first condition fixes the areal radius of the throat or the equator,
and the second
is the extremum condition. The third condition fixes the value of the
chiral profile function at the throat or the equator, which is a free parameter.
In this study we will first focus on the special case $F_0= n \pi/2$,
when the wormhole solutions are symmetric under the interchange
of the asymptotic regions. Subsequently, we will consider non-symmetric
wormhole solutions.

In the asymptotic regions
we impose the following boundary conditions:
\begin{equation}
A(\eta\to \infty) \to 1 \ ,  \ \ \ 
F(\eta\to \infty) \to 0 \ ,  \ \ \ 
F(\eta\to -\infty) \to n \pi \ .
\label{bcasym}
\end{equation}
The first condition sets the time scale, whereas the 
second and third conditions result from requiring finite energy
and topological charge $n$.

\subsection{Topological Charge}

In order to identify the topological charge of the solutions,
we consider the chiral matrix $U$ as a map of spatial slices of the
wormhole space-time to the group manifold $SU(2) \sim S^3$. Since $U$ takes
constant values in both asymptotic regions we can contract each of these 
regions to a point. The spatial slices then become topologically equivalent
to a three dimensional sphere, where the north and south pole correspond to
the asymptotic regions $\eta\to +\infty$ and $\eta\to -\infty$, respectively.
Thus the chiral matrix can be regarded as a map between two three-spheres, and
the topological charge is defined as the degree of the map. 
With the Ansatz for the chiral matrix Eq.~(\ref{skyrmU}) and the above asymptotic boundary 
conditions for the profile function $F$ with $n=1$ the topological charge is equal to one.

\section{Symmetric Wormholes}

In this section we focus on wormholes  that are symmetric under the interchange
of the asymptotic regions.
For the numerical calculations, we re-parametrize the function $R$ as
$R^2 = (\eta^2 +r_0^2) h(\eta)$, since the new function $h$ is bounded 
in the full interval
$-\infty < \eta < \infty$, in contrast to $R$. Then, the boundary conditions
$R(0)=r_0$ and $R'(0)=0$ translate into $h(0)=1$ and  $h'(0)=0$, respectively.

To obtain wormholes symmetric with respect to the throat
(or beyond $\alpha_{\rm cr}$ to the maximum), i.e. $\eta=0$, 
we require for the chiral field that $\cos F$ is an odd function in $\eta$.
This yields $F(0)=F_0 = \pi/2$ when $n=1$. 

The mass $M$ of the solutions is obtained from the asymptotic behaviour of the 
metric function $A$,
\begin{equation}
A \to 1 - \frac{\mu}{\eta} \ , 
\label{Ainf}
\end{equation}
where the dimensionless mass parameter $\mu$ is related to $M$ by $\mu=\alpha M/M_0$,
with $M_0 = 4\pi \kappa$.

The ODEs are solved numerically for the given set of
boundary conditions and parameters $\alpha$ and $r_0$.
The quality of the numerical solutions is good,
since the variation of the constant $D$ as computed from Eq.~(\ref{eqd}) 
is typically less than $10^{-9}$.

\subsection{Non-linear Sigma Model (NLS) Wormholes}

\begin{figure}[t!]
\begin{center}
\vspace{0.5cm}
\mbox{\hspace{-0.5cm}
\subfigure[][]{\hspace{-1.0cm}
\includegraphics[height=.25\textheight, angle =0]{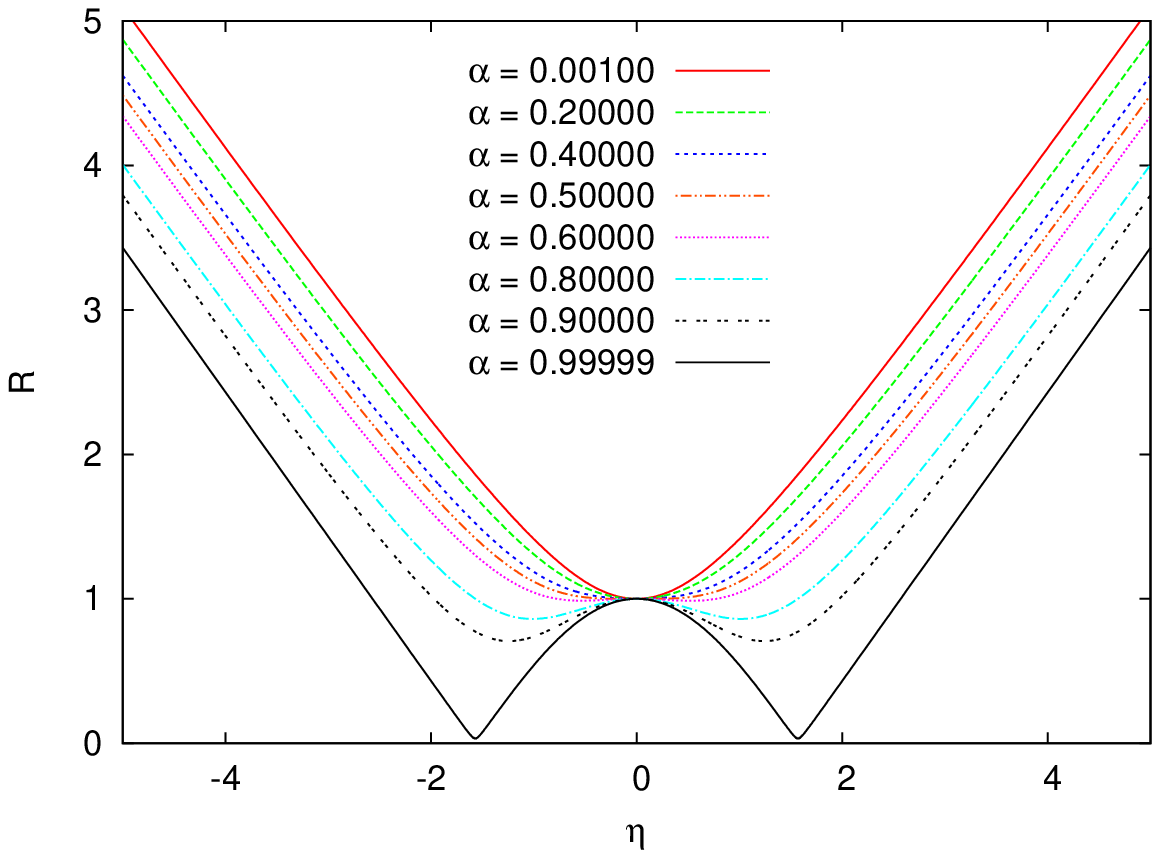}
\label{fig1a}
}
\subfigure[][]{\hspace{-0.5cm}
\includegraphics[height=.25\textheight, angle =0]{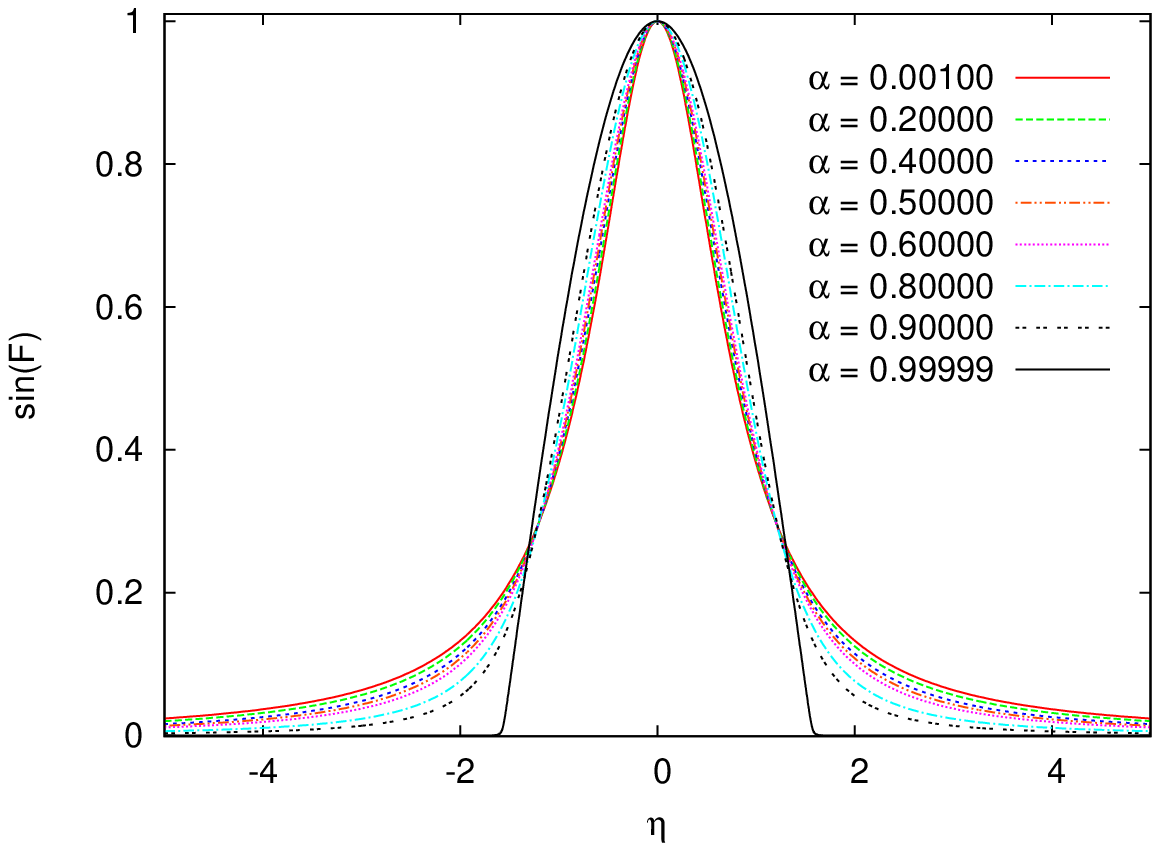}
\label{fig1b}
}
}
\end{center}
\vspace{-0.7cm}
\caption{
Plots of (a) the metric function $R$  and (b) the chiral function
$\sin F$  for several values of $\alpha$ for non-linear sigma model wormholes.
\label{fig1}
}
\end{figure}

We first consider wormholes in the absence of the Skyrme term,
i.e.~we take the limit $1/e^2 \to 0$. In this case the set of Einstein and matter equations 
is invariant under the scalings $\eta \to \lambda \eta$ and $R \to \lambda R$.
In order to fix the scale we choose $r_0=1$. This leaves only $\alpha$ as a free
parameter.

Observe that Eq.~(\ref{eqA}) reduces to $R^2 A'=const$. However, for symmetric
solutions $A'(0)=0$. Thus this constant has to vanish. Consequently, $A=1$ is the only
solution which satisfies the boundary condition $A(\eta\to \infty) \to 1$, which implies that the mass vanishes.

We have solved the coupled set of Einstein-matter equations numerically for $0\leq \alpha < 1$.
As examples we show in Fig.~\ref{fig1} the functions $R(\eta)$ and $\sin F(\eta)$
for several values of $\alpha$. In Fig.~\ref{fig1a} we see that the areal radius
$R$ possesses a minimum at $\eta=0$ when $\alpha \leq \alpha_{\rm cr} = 1/2$.
These solutions possess a single throat.
For larger values of $\alpha$, however, $R$ possesses a local maximum at $\eta=0$ 
and two minima located symmetrically to each side of the local maximum. 
Consequently, these wormholes possess a double throat.
As $\alpha$
is increased the minima become more pronounced.
Thus the size of the two throats shrinks and tends to zero as $\alpha$ tends
to the value one. In this limit the locations of the throats are at 
the values $\pm \pi/2$. The behaviour of the chiral profile function is demonstrated
in Fig.~\ref{fig1b} where we plot $\sin F$ as a function of $\eta$. Note
that the function becomes more and more concentrated on the interval 
$-\pi/2 \leq \eta \leq \pi/2$ as $\alpha$ approaches one.

\begin{figure}[t!]
\begin{center}
\vspace{0.5cm}
\mbox{\hspace{-0.5cm}
\subfigure[][]{\hspace{-1.0cm}
\includegraphics[height=.25\textheight, angle =0]{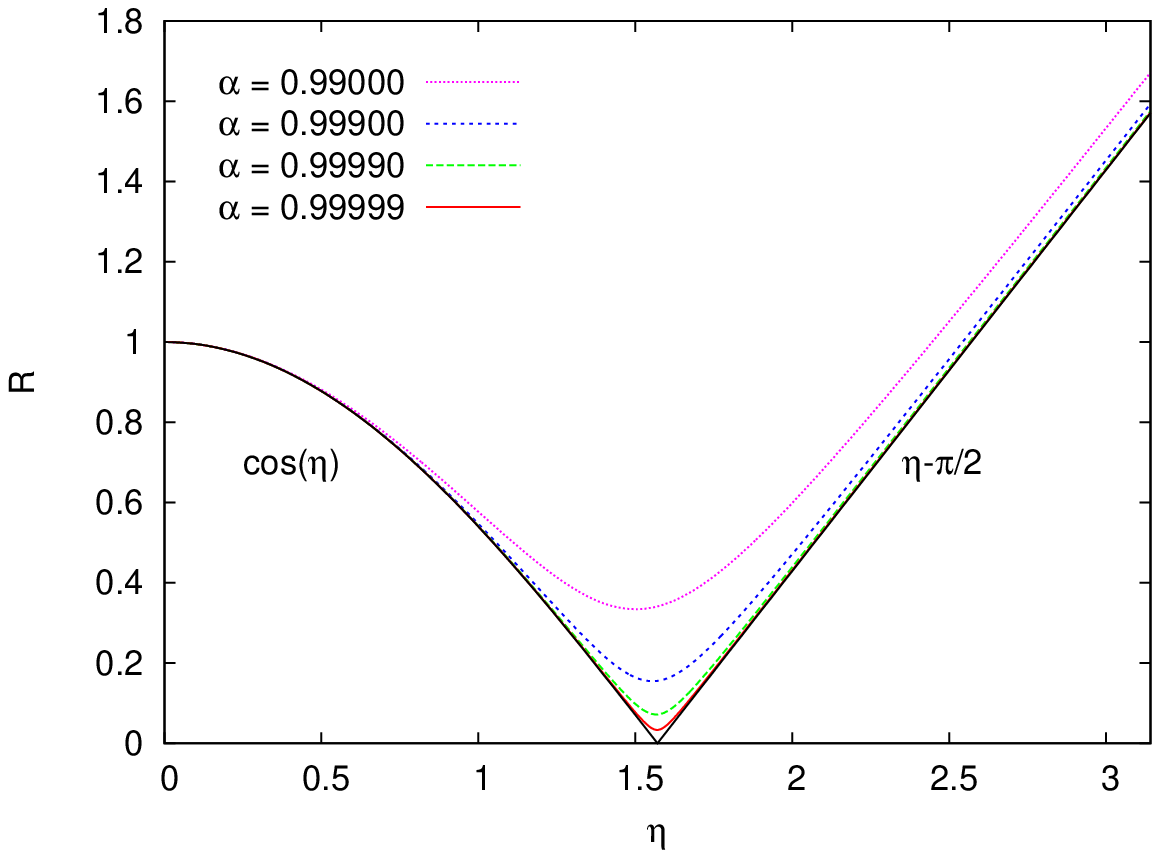}
\label{fig2a}
}
\subfigure[][]{\hspace{-0.5cm}
\includegraphics[height=.25\textheight, angle =0]{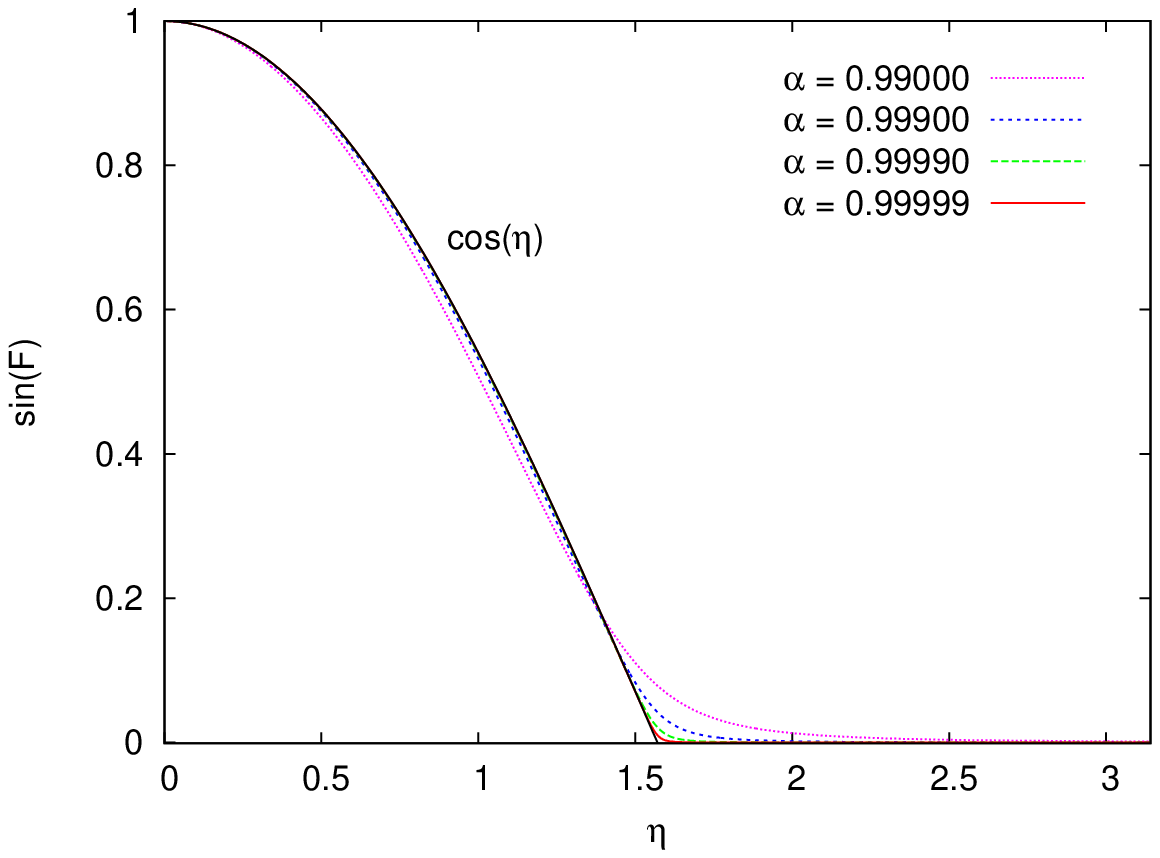}
\label{fig2b}
}
}
\end{center}
\vspace{-0.7cm}
\caption{
Same as Fig. 1 but for  values of $\alpha$ close
to the limit $\alpha=1$.
\label{fig2}
}
\end{figure}

\begin{figure}[h!]
\begin{center}
\vspace{0.5cm}
\mbox{\hspace{-0.5cm}
\subfigure[][]{\hspace{-1.0cm}
\includegraphics[height=.25\textheight, angle =0]{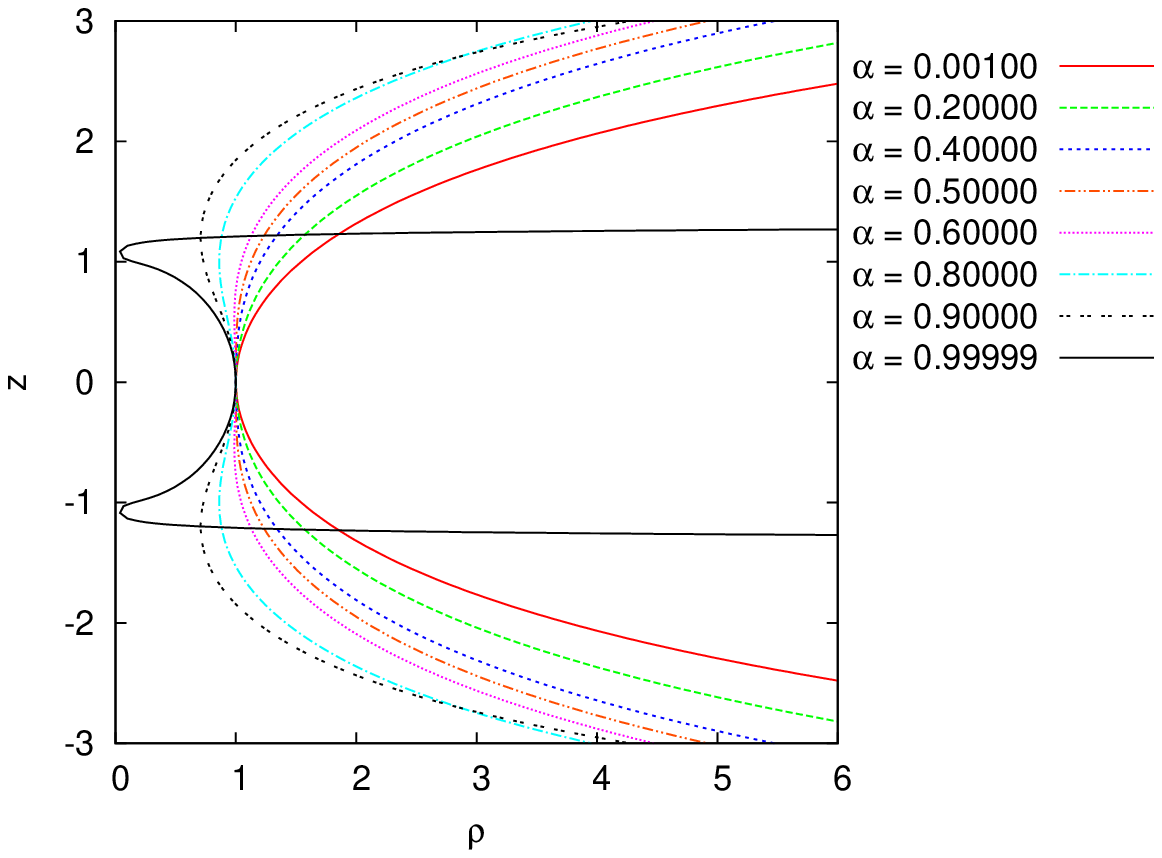}
\label{fig3a}
}
}
\mbox{\hspace{-0.5cm}
\subfigure[][]{\hspace{-0.5cm}
\includegraphics[width=.33\textwidth, angle =0]{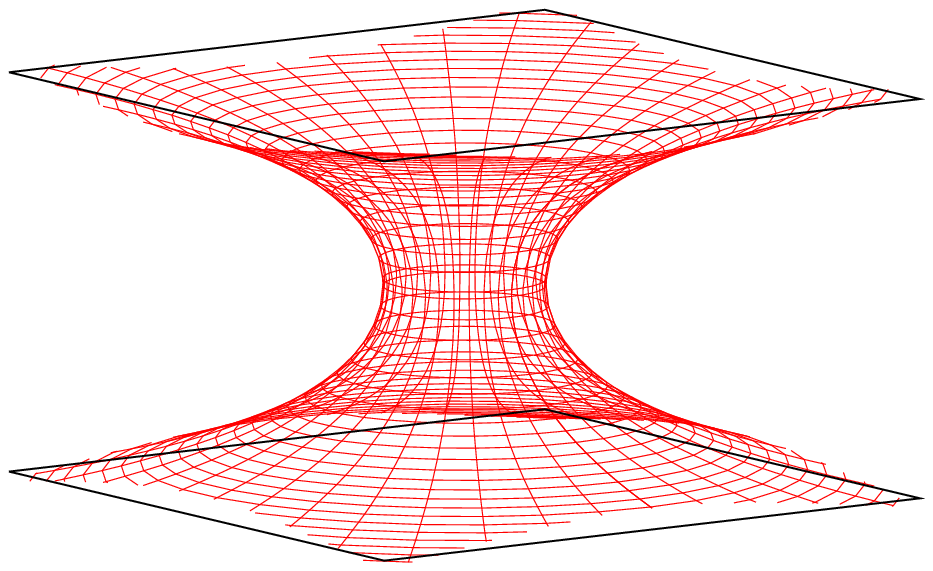}
\label{fig3b}
}
\subfigure[][]{\hspace{-1.0cm}
\includegraphics[width=.33\textwidth, angle =0]{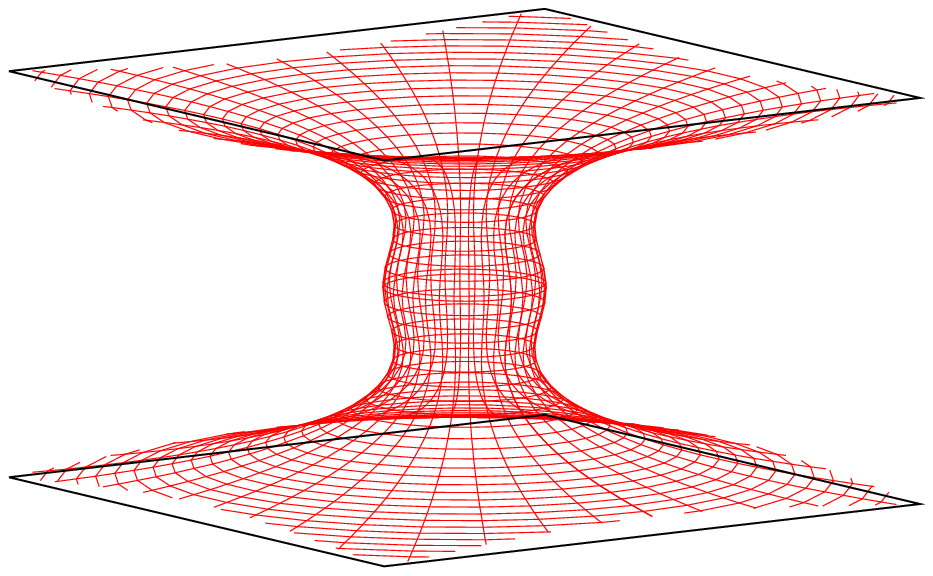}
\label{fig3c}
}
\subfigure[][]{\hspace{-0.5cm}
\includegraphics[width=.33\textwidth, angle =0]{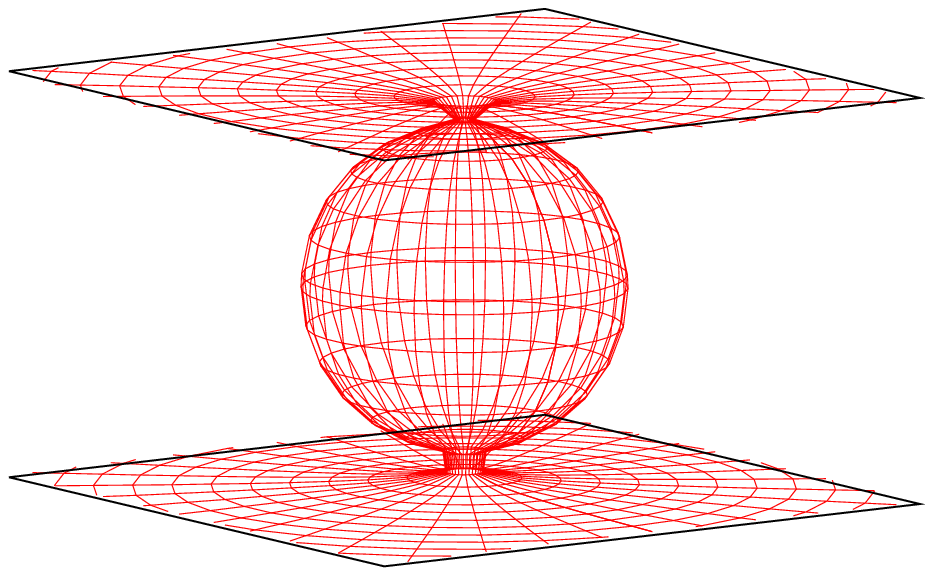}
\label{fig3d}
}
}
\end{center}
\vspace{-0.7cm}
\caption{
(a) The isometric embedding of NLS model wormholes is shown for several values of
$\alpha=0$.
(b-d) Three dimensional view of the isometric embedding 
for (b) $\alpha=0.001$, (c)  $\alpha=0.8$ and  (d) $\alpha=0.99999$.
\label{fig3}
}
\end{figure}

The limit $\alpha \to 1$ is demonstrated in Fig.~\ref{fig2}.
We conclude that the limiting solution consists of three parts.
On the interval $-\pi/2 < \eta < \pi/2$ the areal radius and the 
chiral profile function are
\begin{equation}
R(\eta) = \cos\eta \ , \ \ \ \ F(\eta)= \frac{\pi}{2}-\eta \ .
\label{innersol}
\end{equation}
In this region the metric reads 
\begin{equation}
ds^2 = -dt^2 +d\eta^2 + \cos^2\eta \, d\Omega_2^2 = -dt^2 + d\Omega_3^2 \ , 
\end{equation}
and describes an Einstein universe, ${\cal M}_E = R_t \times S^3$.
The chiral matrix $U$ is a one-to-one mapping of the three sphere $S^3$ to $SU(2) \sim S^3$.
It can be easily  checked that the system ~(\ref{innersol}) together with $A=1$ is indeed
an exact solution of the Einstein-matter equations on the interval 
$-\pi/2 < \eta < \pi/2$ \cite{Lechner:2000bw}. Moreover, the scalar charge $D$ vanishes for this solution
as can be seen from Eq.~(\ref{eqd}). This implies that there is no exotic matter 
present.

In the outer regions the solution reads
\begin{eqnarray}
& &
R(\eta) =\ \ \eta-\frac{\pi}{2} \ , \ \ \ \ F(\eta)=0 \  \ \ (\eta>\frac{\pi}{2}),
\\
& &
R(\eta) =  -\eta+\frac{\pi}{2} \ , \ \ \ \ F(\eta)=\pi \  \ \ (\eta<-\frac{\pi}{2}).
\end{eqnarray}
In both regions the metric describes the Minkowski space-time,
\begin{equation}
ds^2 = -dt^2 +dR^2 + R^2d\Omega_2^2 \ , 
\end{equation}
and the chiral matrix $U$ is the trivial map $U=\pm 1$. Also in these regions the 
scalar charge vanishes, as it should for a vacuum solution.

The resulting space-time consists of two distinct Minkowski space-times plus
one Einstein universe. The origin of one of the Minkowski space-times is identified 
with the  north pole of the spatial $S^3$ of the Einstein universe, 
and the origin of the other Minkowski space-time is identified with the south pole
of the $S^3$.

Note, that at the ``glueing points" (i.e. $\eta=\pm \pi/2$) 
the derivative of the function $R$ and the profile function $F$ possess jump
discontinuities which lead to $\delta$-function terms in the 
second derivatives. However, the metric depends on the function $R^2$,
which has a discontinuity only in the fourth derivative. 
Hence the Ricci tensor and the Kretschmann scalar possess discontinuities 
at $\eta=\pm \pi/2$, but no $\delta$-function terms
\footnote{
The analytic solution corresponds to the regular part of the limiting solution.
In the limit $\alpha \to 1$ in the Ricci and the Kretschmann scalar
$\delta$-function singularities develop at the points $\eta = \pm \pi/2$.
}.

The geometry of a spatial hyper-surface of the wormhole space-times is visualized
in Fig.~\ref{fig3}. Here we show examples of the isometric embedding of the equatorial plane
$\theta= \pi/2$ for several values of $\alpha$. 
The embedding is given by the parametric representation
\begin{equation}
\rho(\eta) = R(\eta) \ , \ \ \ \ 
z(\eta)= \int_0^\eta\sqrt{1-R'^2} \ d\eta' \ .
\label{embedd}
\end{equation}
We observe that the radius $R$ has a single minimum at $z=0$ if 
$\alpha< \alpha_{\rm cr}$, corresponding to the waist in Fig.~\ref{fig3b}. The 
minimum becomes degenerate if $\alpha=\alpha_{\rm cr}$ and turns to a (local) 
maximum for larger values of $\alpha$, corresponding to the belly in Fig.~\ref{fig3c}.
Close to the limit $\alpha\to 1$ the outer regions are represented by the two plains
whereas the inner region forms a sphere, as shown in Fig.~\ref{fig3d}.


\subsection{Skyrmionic Wormholes} 

We now turn to wormholes in the presence of the Skyrme term.
In this case the coupling parameter $e$ can be included in the scaled
quantities by letting
\begin{equation}
\eta = \frac{\tilde{\eta}}{e\kappa} \ , \ \ \ 
R=\frac{\tilde{R}}{e\kappa} \ , \ \ \ 
\phi = \kappa \tilde{\phi} \ , \ \ \ 
4 \pi G = \frac{\alpha}{\kappa^2}\ , \ \ \  
D = \tilde{D}e \ , 
\label{scalesk}
\end{equation}
or, equivalently, by setting $e=1$.

\begin{figure}[h!]
\begin{center}
\vspace{0.5cm}
\mbox{\hspace{-0.5cm}
\subfigure[][]{\hspace{-1.0cm}
\includegraphics[height=.25\textheight, angle =0]{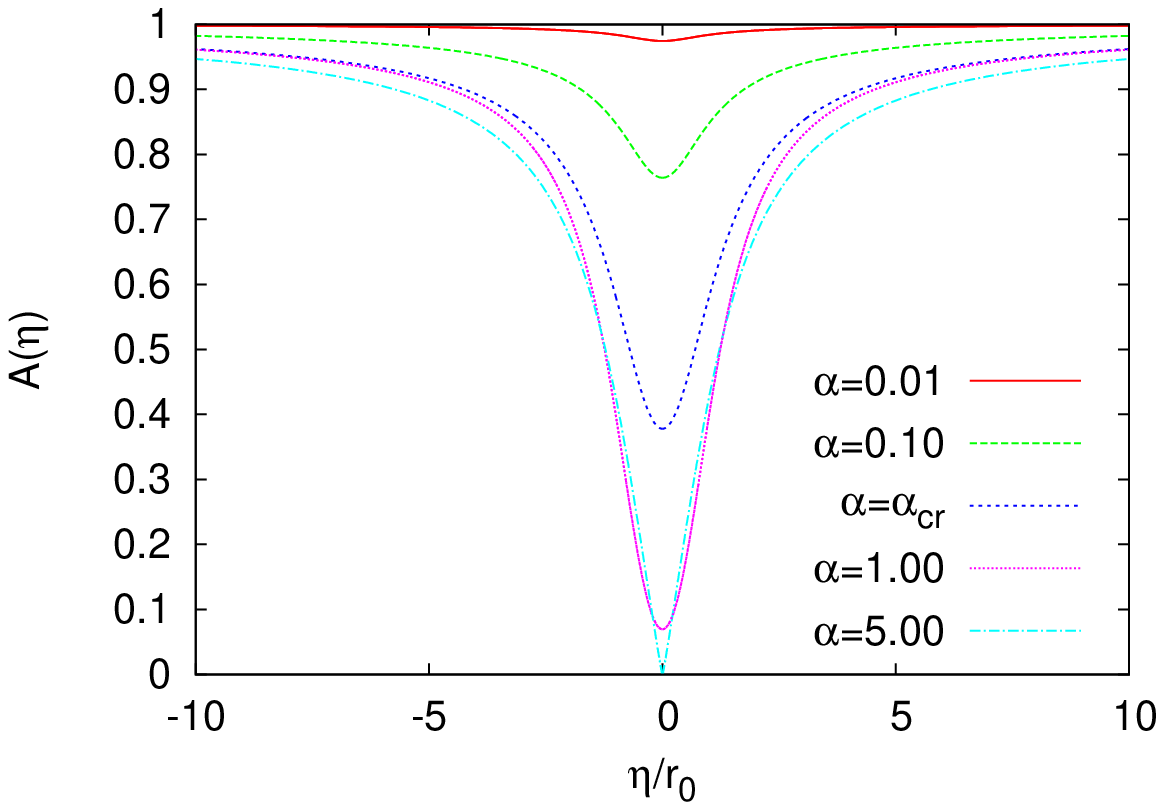}
\label{fig4a}
}
\subfigure[][]{\hspace{-0.5cm}
\includegraphics[height=.25\textheight, angle =0]{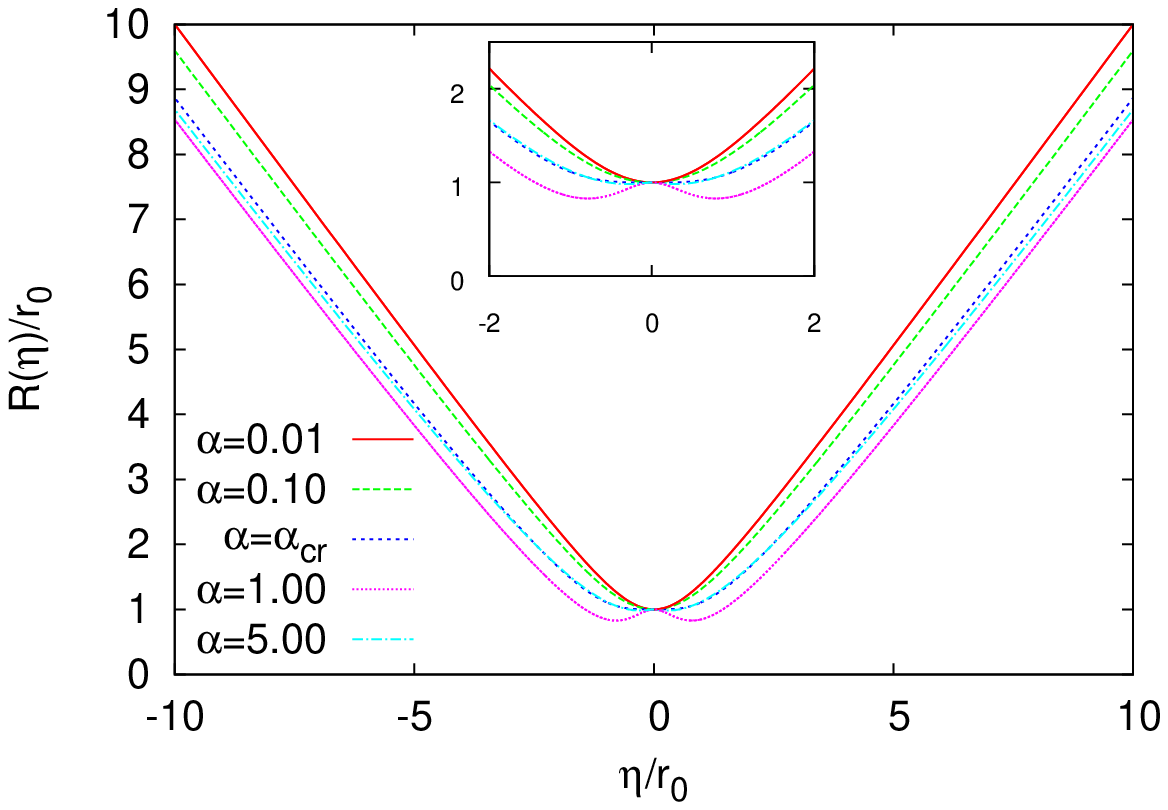}
\label{fig4b}
}
}
\mbox{\hspace{-0.5cm}
\subfigure[][]{\hspace{-1.0cm}
\includegraphics[height=.25\textheight, angle =0]{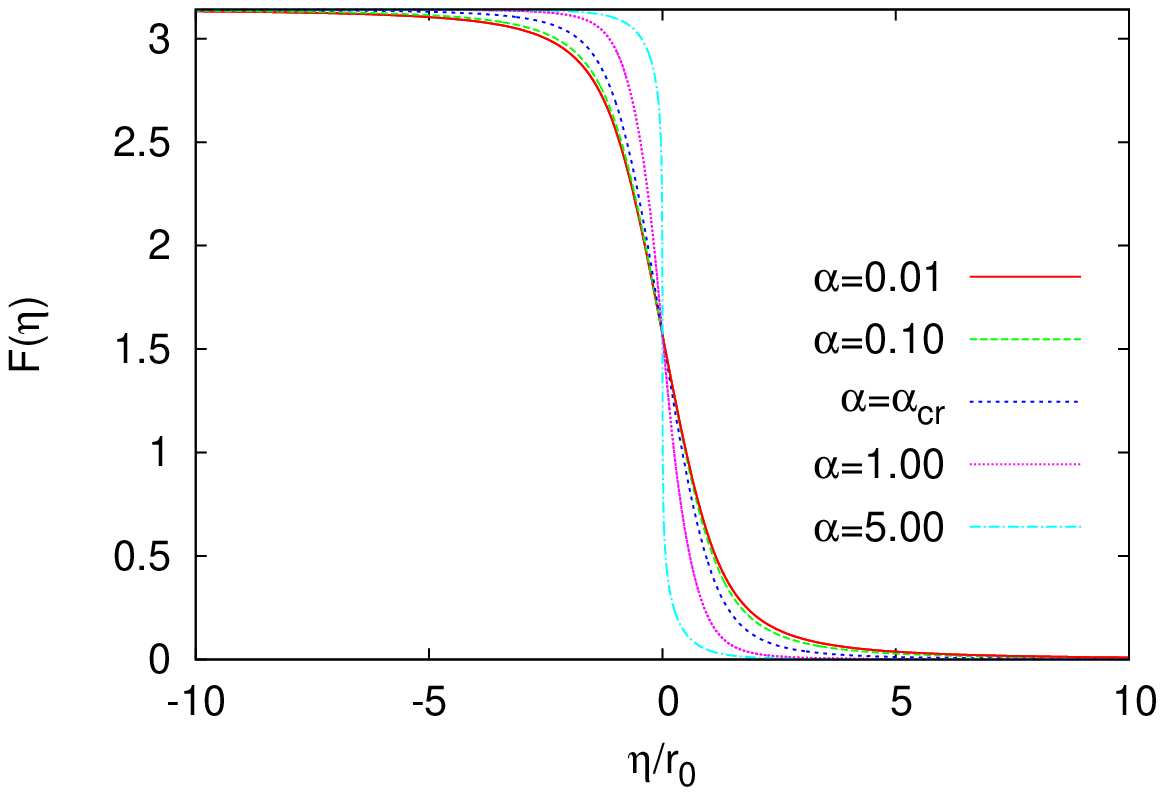}
\label{fig4c}
}
\subfigure[][]{\hspace{-0.5cm}
\includegraphics[height=.25\textheight, angle =0]{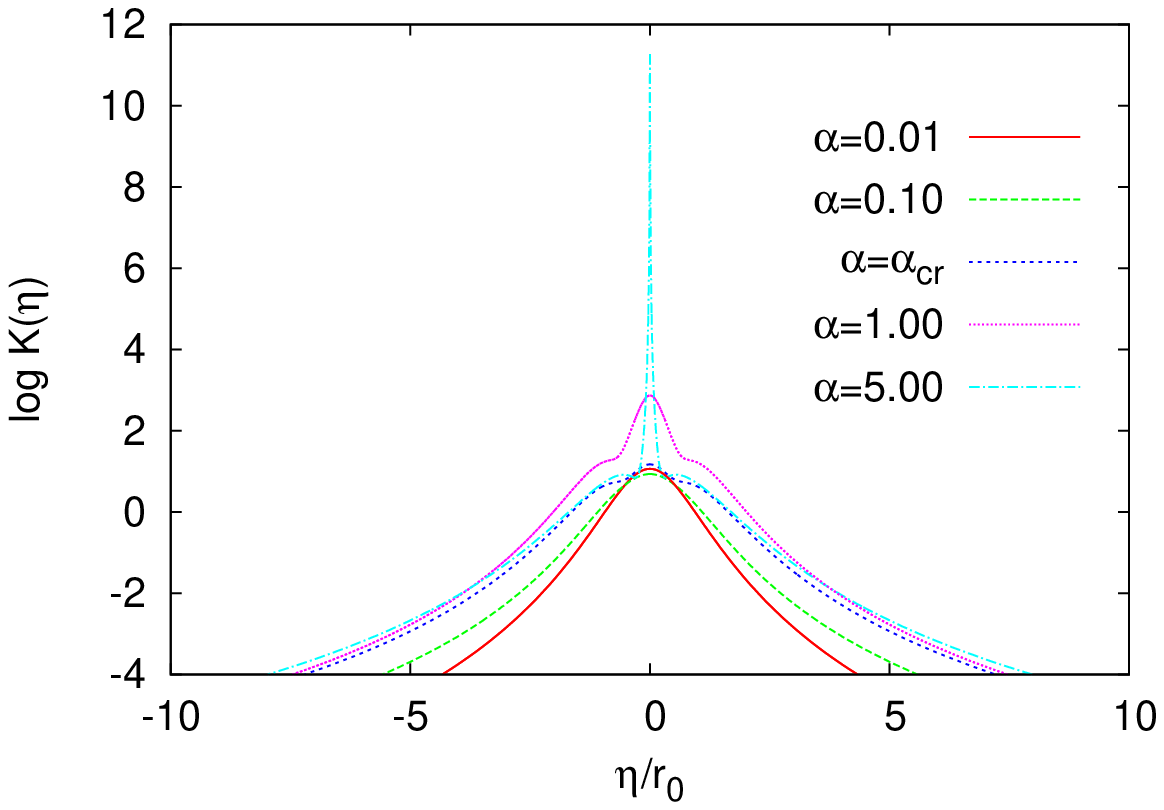}
\label{fig4d}
}
}
\end{center}
\vspace{-0.7cm}
\caption{
Plots of (a) the metric function $A$, (b)
the function $R$, (c)
the chiral profile function $F$  and (d)
the Kretschmann scalar $K$ 
are presented for Skyrmionic wormholes when $r_0=1$, 
$\alpha=0.01$, $0.1$  and $\alpha_{\rm cr}=1/3$, $1.0$, $5.0$. 
\label{fig4}
}
\end{figure}

If the Skyrme term is present, the corresponding Einstein-matter equations are
{\it no longer scale-invariant}. Then  $r_0$ and $\alpha$ can be considered 
as free parameters.
In Fig.~\ref{fig4} we show examples of solutions for $r_0=1$ for different 
values of $\alpha$.
We observe that in the limit of vanishing $\alpha$ the metric functions approach
the massless Ellis solution, that is, $A(\eta)=1$ and $R^2(\eta) = \eta^2+r_0^2$.
Thus, we obtain a Skyrmion in the background of the Ellis wormhole
space-time in this limit. As $\alpha$ is increased from zero the metric functions 
deviate from the Ellis solution. From Fig.~\ref{fig4b} we see that 
when $\alpha$ exceeds a critical value, the
minimum of the areal radius splits into two minima separated by a local maximum.
Thus again the single throat splits into a double throat and an equator inbetween.
As $\alpha$ becomes large the metric
function $A$ tends to zero at the equator, while the Kretschmann scalar tends
to diverge at the equator.

\begin{figure}[h!]
\begin{center}
\vspace{0.5cm}
\mbox{\hspace{-0.5cm}
\subfigure[][]{\hspace{-1.0cm}
\includegraphics[height=.25\textheight, angle =0]{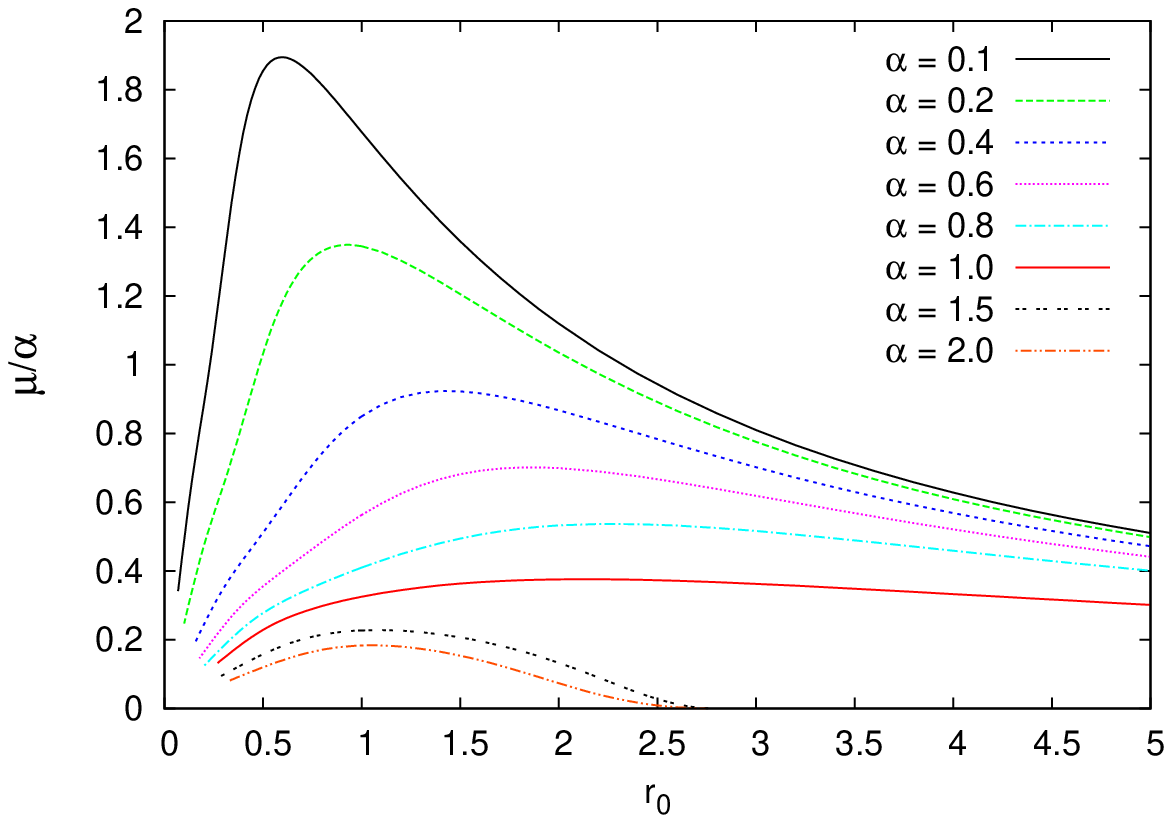}
\label{fig5a}
}
\subfigure[][]{\hspace{-0.5cm}
\includegraphics[height=.25\textheight, angle =0]{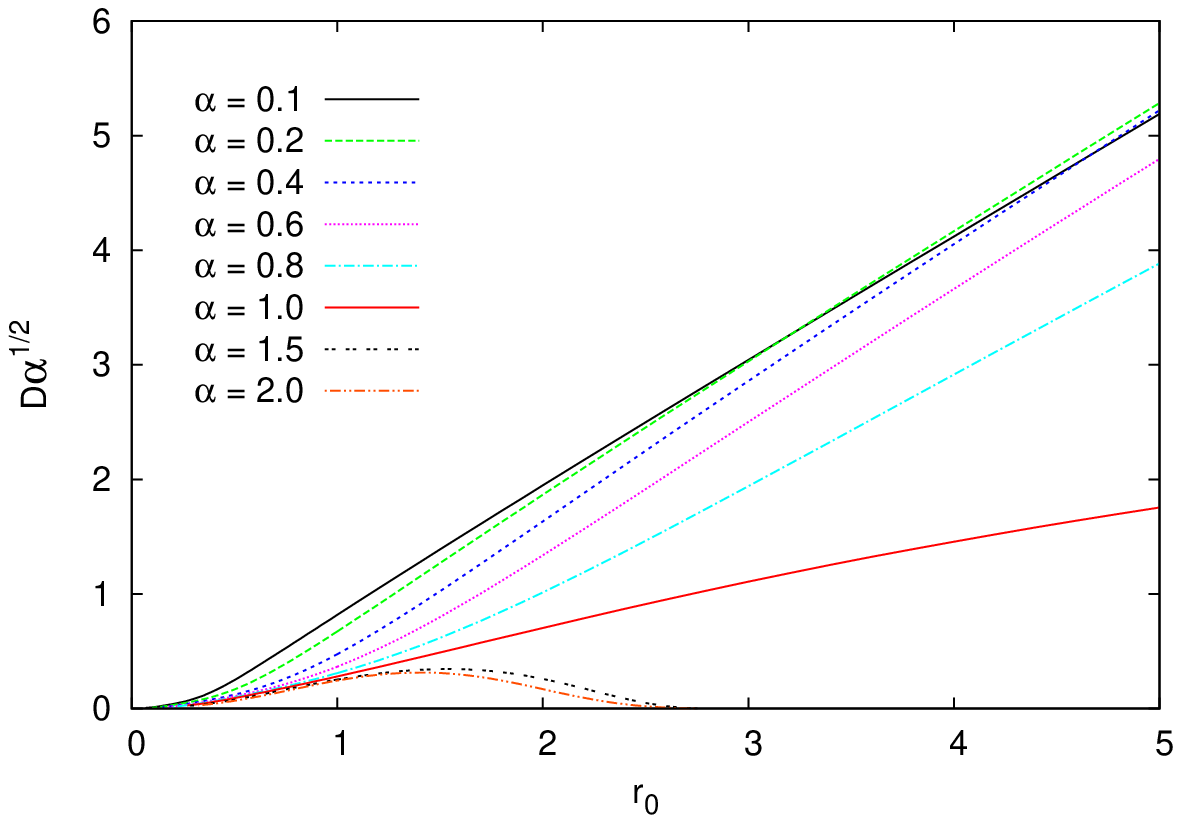}
\label{fig5b}
}
}
\end{center}
\vspace{-0.7cm}
\caption{
Plots of (a) the  scaled mass $\mu/\alpha$  and (b)  the scaled scalar charge $D\sqrt{\alpha}$ 
are shown for Skyrmionic wormholes as functions of $r_0$ for
different values of $\alpha$.
\label{fig5}
}
\end{figure}

In Figs.~\ref{fig5a} and \ref{fig5b} plots of  the scaled mass 
$\mu/\alpha = M/M_0$ with $M_0 = 4\pi \kappa/e$ and the scaled
scalar charge $\sqrt{\alpha}D$ are presented versus $r_0$ for different values of 
the parameter $\alpha$.
We observe that for $\alpha \leq 1$ solutions exist for arbitrarily large values 
of $r_0$. In order to identify the limit $r_0 \to \infty$ we introduce the scaled
coordinate $\hat{\eta} = \eta/r_0$ and the scaled function $\hat{R} = R/r_0$. From 
the Einstein equations (\ref{eqR}), (\ref{eqA}) and the chiral equation (\ref{eqSk})
it follows that in the limit $r_0 \to \infty$ all terms resulting from the Skyrme term
are suppressed as $1/r_0^2$ and will vanish. However, this is exactly the 
limit $1/e^2 \to 0$ discussed in the previous section. 
Thus, the Skyrmionic wormholes  approach the NLS model wormholes 
(after rescaling)
when $r_0$ becomes very large, provided that $\alpha \leq 1$.

Let us now consider the case $\alpha > 1$.
In this case,  NLS  wormholes {\it do not} exist any more and we expect a
different scenario. Indeed, we see from 
Figs.~\ref{fig5a} and \ref{fig5b} that Skyrmionic wormholes exist only up to a maximal
value $r_{0,\rm max}$, which depends on $\alpha$. The mass and the scalar charge 
tend to zero when $r_0$ tends to $r_{0,\rm max}$.

\begin{figure}[h!]
\begin{center}
\vspace*{0.5cm}
\mbox{\hspace{-0.5cm}
\subfigure[][]{\hspace{-1.0cm}
\includegraphics[height=.25\textheight, angle =0]{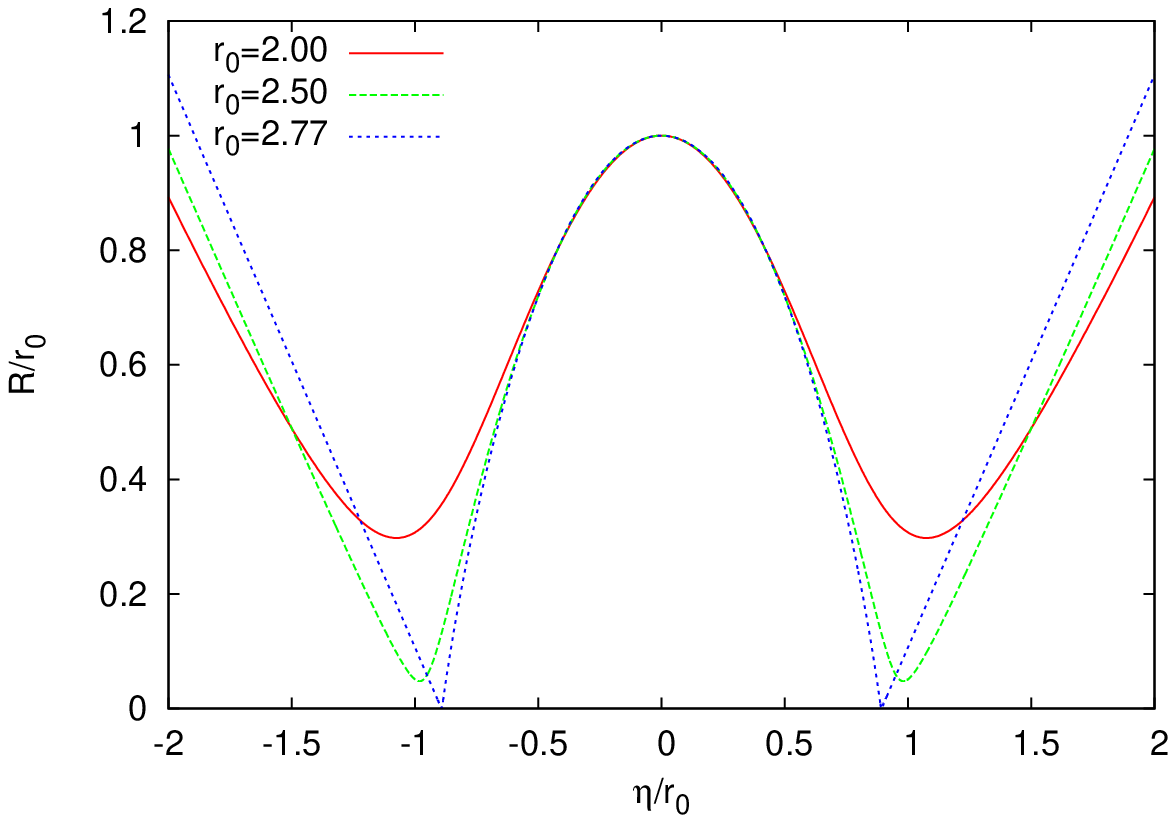}
\label{fig6a}
}
\subfigure[][]{\hspace{-0.5cm}
\includegraphics[height=.25\textheight, angle =0]{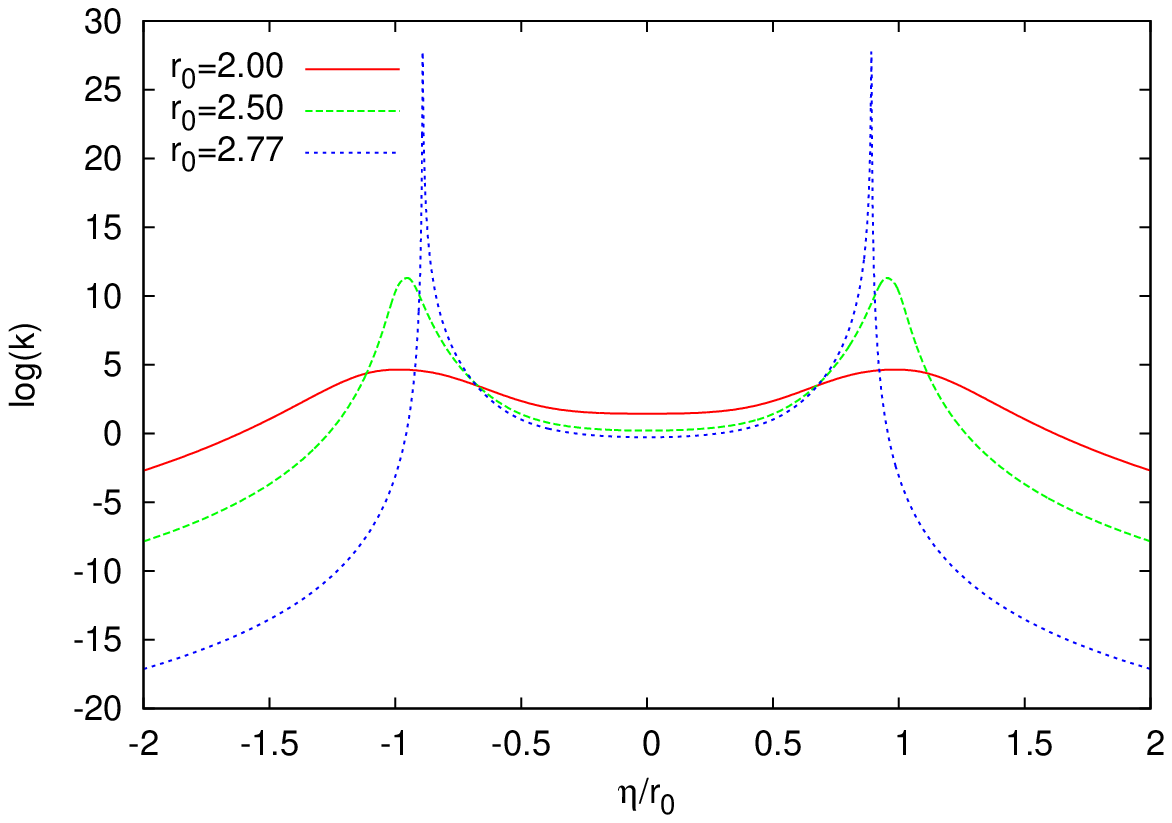}
\label{fig6b}
}
}
\end{center}
\vspace{-0.7cm}
\caption{
Plots of (a) the function $R$ and (b)   the Kretschmann scalar $K$ 
  for Skyrmionic wormholes when $\alpha=1.5$ for different 
 values of $r_0$ close to the maximal value $r_{0,\rm max}$.
\label{fig6}
}
\end{figure}

In Fig.~\ref{fig6} we demonstrate that in the limit 
$r_0 \to r_{0,\rm max}$ the solutions become singular.
In particular, Fig.~\ref{fig6a}  displays the function $R$ when  $\alpha=1.5$ for several values of
$r_0$ approaching the maximal value $r_{0,\rm max}$. 
We see that the minimum of $R$ approaches zero
at the points $\pm \eta_0$. In Fig.~\ref{fig6b} it is seen that the Kretschmann
scalar diverges at $\pm \eta_0$ when $r_0$ tends to the maximal value $r_{0,\rm max}$.

\begin{figure}[t!]
\begin{center}
\vspace{-0.5cm}
\mbox{\hspace{-0.5cm}
\subfigure[][]{\hspace{-1.0cm}
\includegraphics[height=.25\textheight, angle =0]{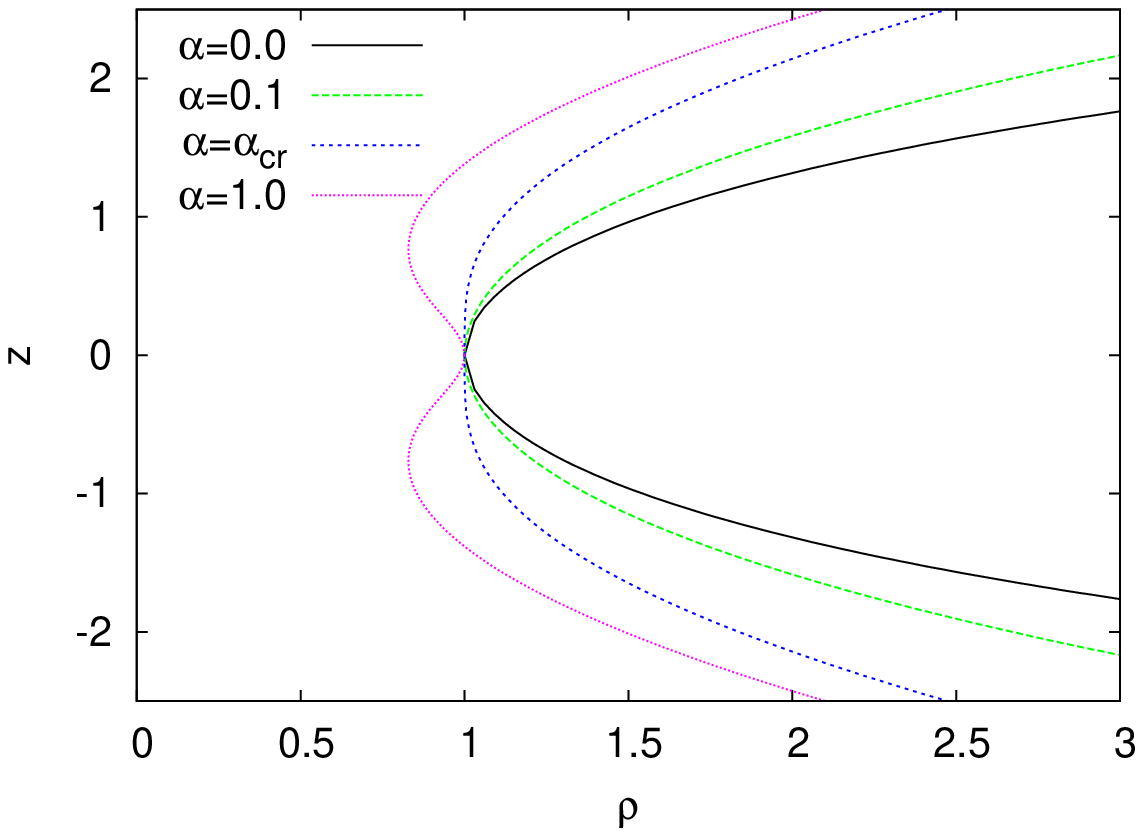}
\label{fig7a}
}
}
\mbox{\hspace{-0.5cm}
\subfigure[][]{\hspace{-0.5cm}
\includegraphics[width=.33\textwidth, angle =0]{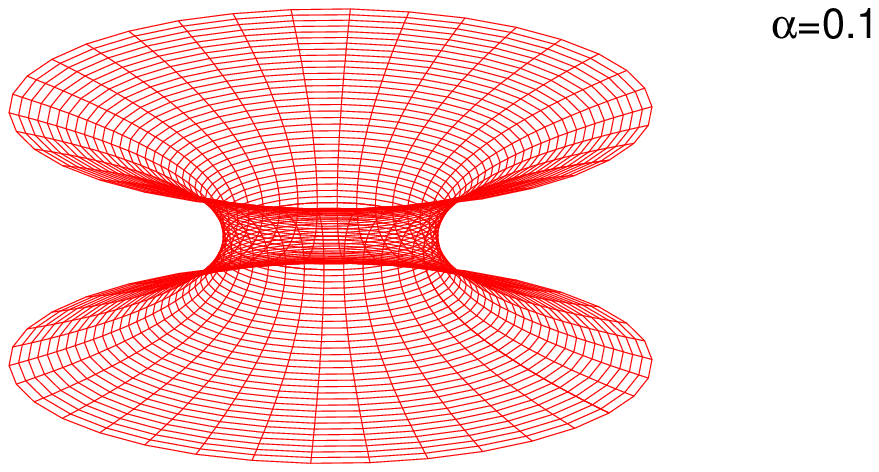}
\label{fig7b}
}
\subfigure[][]{\hspace{-1.0cm}
\includegraphics[width=.33\textwidth, angle =0]{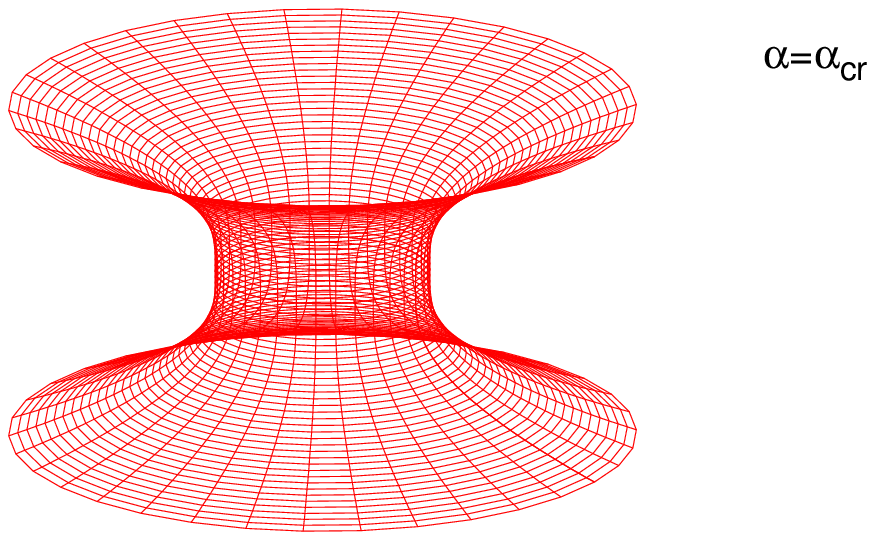}
\label{fig7c}
}
\subfigure[][]{\hspace{-0.5cm}
\includegraphics[width=.33\textwidth, angle =0]{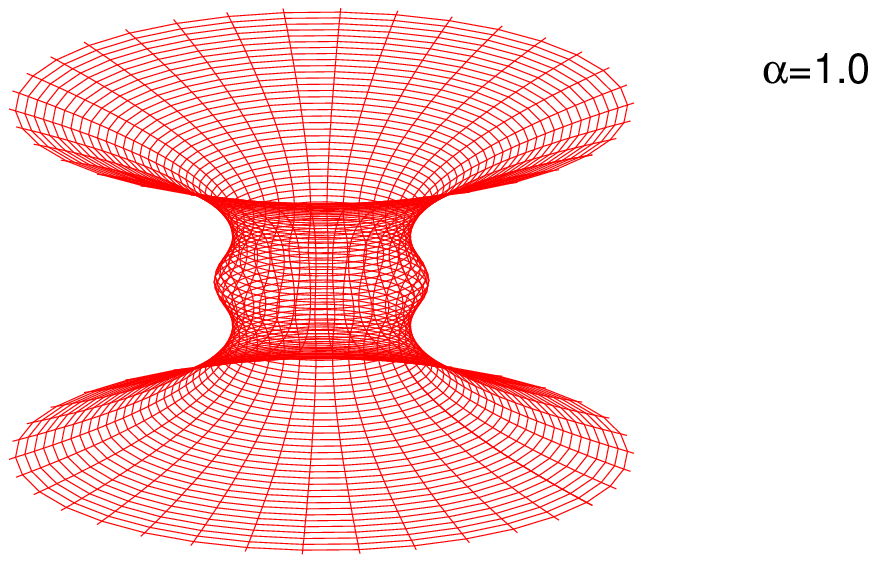}
\label{fig7d}
}
}
\end{center}\vspace{-0.7cm}
\caption{
(a) The isometric embedding of Skyrmionic wormholes is presented when  $r_0=1$ while
$\alpha=0$ (Ellis), $\alpha=0.1$, $\alpha=\alpha_{\rm cr}$ and $\alpha=1.0$.
(b-d) Three dimensional view of the isometric embedding of the wormholes 
when  $r_0=1$ while  (b) $\alpha=0.1$, (c) $\alpha=\alpha_{\rm cr}$ and (d) $\alpha=1.0$.
\label{fig7}
}
\end{figure}

The geometry of a spatial hyper-surface of the wormhole space-times is visualized
in Fig.~\ref{fig7}. Here the isometric embedding of the equatorial plane
$\theta= \pi/2$ for wormholes is presented when $r_0=1$ for several values of $\alpha$
(as examples). We observe that the radius $R$ has a single minimum at $z=0$ if 
$\alpha< \alpha_{\rm cr}$, corresponding to the waist in Fig.~\ref{fig7b}. The 
minimum becomes degenerate at $\alpha=\alpha_{\rm cr}$ and turns into a (local) 
maximum for larger values of $\alpha$, corresponding to the belly presented in Fig.~\ref{fig7d}.

Incidentally, we have also obtained Skyrmionic wormholes
with higher topological charge, but not yet performed a systematic study
of them.

\section{Stability Analysis}

The stability of wormholes is crucial for their physical relevance.
Recently, it was shown that the Ellis wormhole possesses an unstable
mode \cite{Gonzalez:2008wd,Bronnikov:2011if}. 
The unstable mode was missed before because the gauge condition
had been taken too stringent. Here we do not follow the approach 
of \cite{Gonzalez:2008wd,Bronnikov:2011if}, since this
would require the metric in closed form. 
Instead, we make a careful choice of the gauge condition,
which ensures that we do not miss the unstable mode present in the 
Ellis wormhole.
In fact we find that all our chiral wormholes are {\it unstable}.

We restrict to a linear stability analysis in the spherically symmetric sector.
Our starting point is the spherically symmetric metric in the general form
\begin{equation}
ds^2 = -e^{2\nu(t,\eta)} e^{\lambda(t,\eta)}dt^2 + e^{-\lambda(t,\eta)} d\eta^2 
      +e^{\sigma(t,\eta)} d\Omega_2^2 \ ,
\end{equation}
and the time-dependent Ans\"atze for the chiral matrix and the phantom field
\begin{equation}
U = \cos F(t,\eta) + i\sin F(t,\eta) \ \vec{e}\cdot \vec{\tau} \ , 
\ {\rm resp.}\  \ \phi = \phi(t,\eta) \ .
\label{skyrmUt}
\end{equation}

We derive the Einstein-matter equations for these Ans\"atze and consider the 
perturbed metric, scalar field and chiral profile functions
\begin{eqnarray}
\nu(\eta,t)     & = & \bar{\nu}(\eta)     + \delta\nu(\eta)     e^{i\omega t} ,\nonumber \\
\lambda(\eta,t) & = & \bar{\lambda}(\eta) + \delta\lambda(\eta) e^{i\omega t} ,\nonumber \\
\sigma(\eta,t)  & = & \bar{\sigma}(\eta)  + 2\delta\sigma(\eta) e^{i\omega t}, \nonumber \\
\phi(\eta,t)    & = & \bar{\phi}(\eta)    + \delta\phi(\eta)    e^{i\omega t}, \nonumber \\
F(\eta,t)       & = & \bar{F}(\eta)       + \delta F(\eta)      e^{i\omega t}. \nonumber 
\end{eqnarray} 
Here $\bar{F}$ and $\bar{\phi}$ denote the unperturbed chiral and phantom field functions,
respectively, whereas the unperturbed metric functions obey 
\begin{equation}
e^{\bar{\nu}}     = A , \ \ \ 
e^{\bar{\lambda}} = 1 , \ \ \ 
e^{\bar{\sigma}}  = R^2 \ .
\nonumber
\end{equation}
Next we expand the Einstein-matter equations up to first order
in the small quantities 
$\delta\nu(\eta)$, $\delta\lambda(\eta)$, $\delta\sigma(\eta)$, $\delta\phi(\eta)$ 
and $\delta F(\eta)$. The resulting set of linear ODEs for the perturbations form
an eigenvalue problem with eigenvalue $\omega^2$. If  $\omega^2$ is negative
the perturbations increase in time, indicating that the solution is unstable.

We can use the gauge freedom to reduce the number of linear ODEs.
Consider the ODE for the perturbation of the phantom field 
\begin{equation}
\left[ \frac{R^2}{A}\left(\delta\phi' 
+\bar{\phi}'\left(2\delta\sigma +\delta\lambda+\delta\nu\right)\right)\right]'
+\omega^2 \frac{R^2}{A} \delta\phi = 0 \ .
\label{delphieq}
\end{equation}
This form suggests the gauge condition \cite{newpap}
\begin{equation}
\delta\nu + \delta\lambda +2\delta \sigma = 0 \ .
\label{gaugecon}
\end{equation}
Moreover, we argue that, with (\ref{gaugecon}), the function $\delta \phi$ vanishes 
identically, if there exists an unstable mode, i.e. if $\omega^2$ is negative.
Indeed, taking the gauge condition Eq.~(\ref{gaugecon}) into account we multiply
Eq.~(\ref{delphieq}) by $\delta\phi$ and integrate over $(-\infty , \infty)$.
This yields after integration by parts 
\begin{equation}
\left. \left(\frac{R^2}{A} \delta\phi  \delta\phi'\right)\right|_{-\infty}^{\infty}
  = 
     \int_{-\infty}^{\infty}\left[ \frac{R^2}{A}\left({\delta\phi'}^2 
                                     -\omega^2 \delta\phi^2 \right) \right]d\eta\ .
\nonumber
\end{equation}
Since the left-hand-side of this equation  vanishes for normalizable $\delta\phi$, the integral on the right-hand-side  also
vanishes. However, for negative $\omega^2$ the integrand is positive. Therefore, the integral
can only vanish if $\delta\phi$ is identically zero.

With $\delta\phi=0$ and $\delta\nu = -\delta\lambda - 2 \delta\sigma$ 
the resulting equations consist of three second order ODEs for the functions 
$\delta\sigma$, $\delta\lambda$ and $\delta F$ in addition to the constraint
\begin{equation}
R' \delta\lambda + 
2 A\left[\frac{R}{A}\delta\sigma\right]'
+ 2\alpha \frac{\bar{F}'\delta F}{R} \left( \frac{2 \sin^2 F}{e^2} + R^2\right) = 0\ .
\label{constrdellam}
\end{equation}
In principle,  the constraint for the function 
$\delta\lambda$ can be solved in order to reduce 
the three second order ODEs to only two for the corresponding  functions $\delta\sigma$ 
and $\delta F$.
However, this would introduce factors $1/R'$ in the ODEs, which diverge when $R$ becomes extremal. 
Therefore, we choose to solve the system of the three ODEs numerically and just verify
 that the constraint Eq.~(\ref{constrdellam}) is satisfied.

The boundary conditions follow from the requirement that the 
perturbations have to vanish in the asymptotic regions, that is, 
\begin{equation}
 \delta\sigma(\pm\infty)= \delta\lambda(\pm\infty) =\delta F(\pm\infty)=0 \ .
\label{bcpert}
\end{equation}
In addition, in order   to ensure that the perturbations are 
normalizable we impose the condition $\sigma(0)=1$. The eigenvalue $\omega^2$ has to be adjusted such that the perturbations 
satisfy the asymptotic boundary 
conditions \footnote{Technically we define an auxiliary function  $q =\omega^2$ and add the 
ODE: $q'=0$ to the systems of ODEs, without imposing a boundary condition for $q$.
Then, the number of boundary conditions of the Eqs. in (\ref{bcpert}) matches the total order of the
system of ODEs. The value of $q$ is  computed together with the solutions.}.

%
%
 
\begin{figure}[h!]
\begin{center}
\vspace{0.5cm}
\mbox{\hspace{-0.5cm}
\subfigure[][]{\hspace{-1.0cm}
\includegraphics[height=.27\textheight, angle =0]{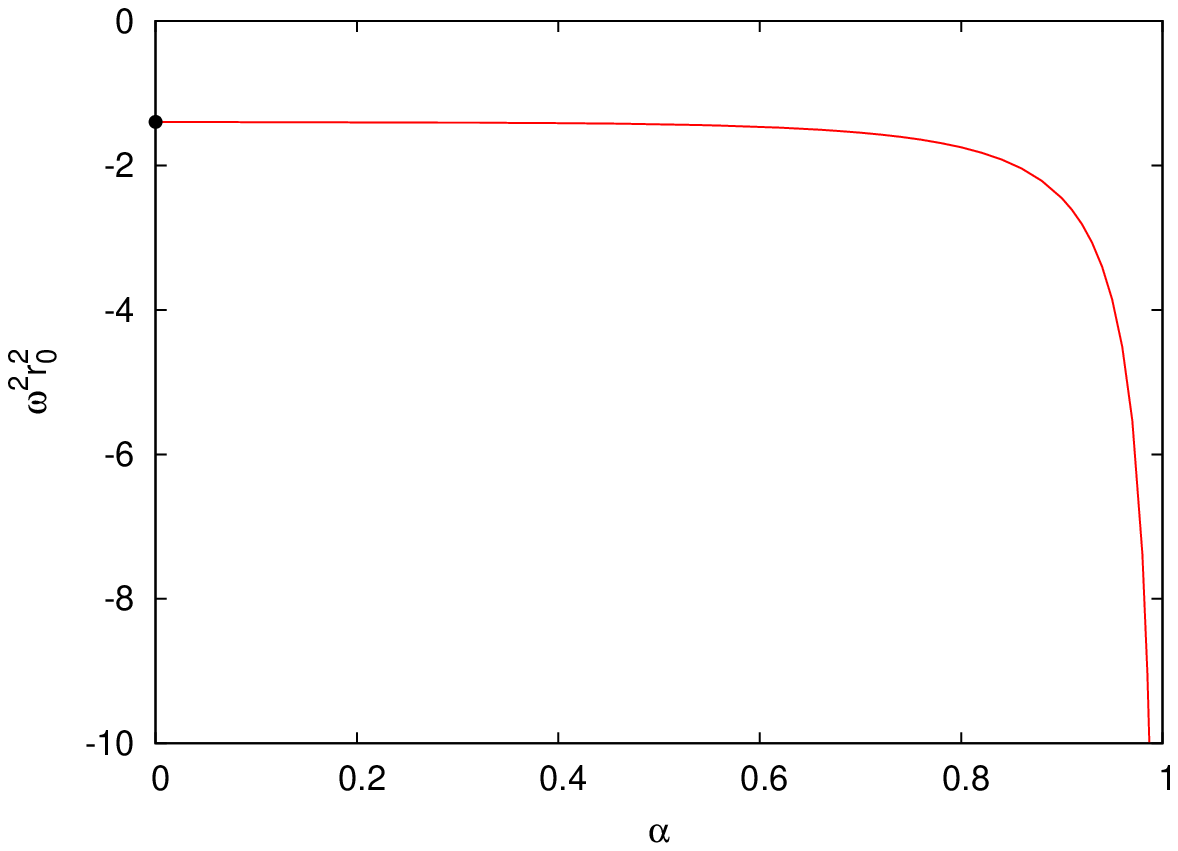}
\label{fig8a}
}
\subfigure[][]{\hspace{-0.5cm}
\includegraphics[height=.27\textheight, angle =0]{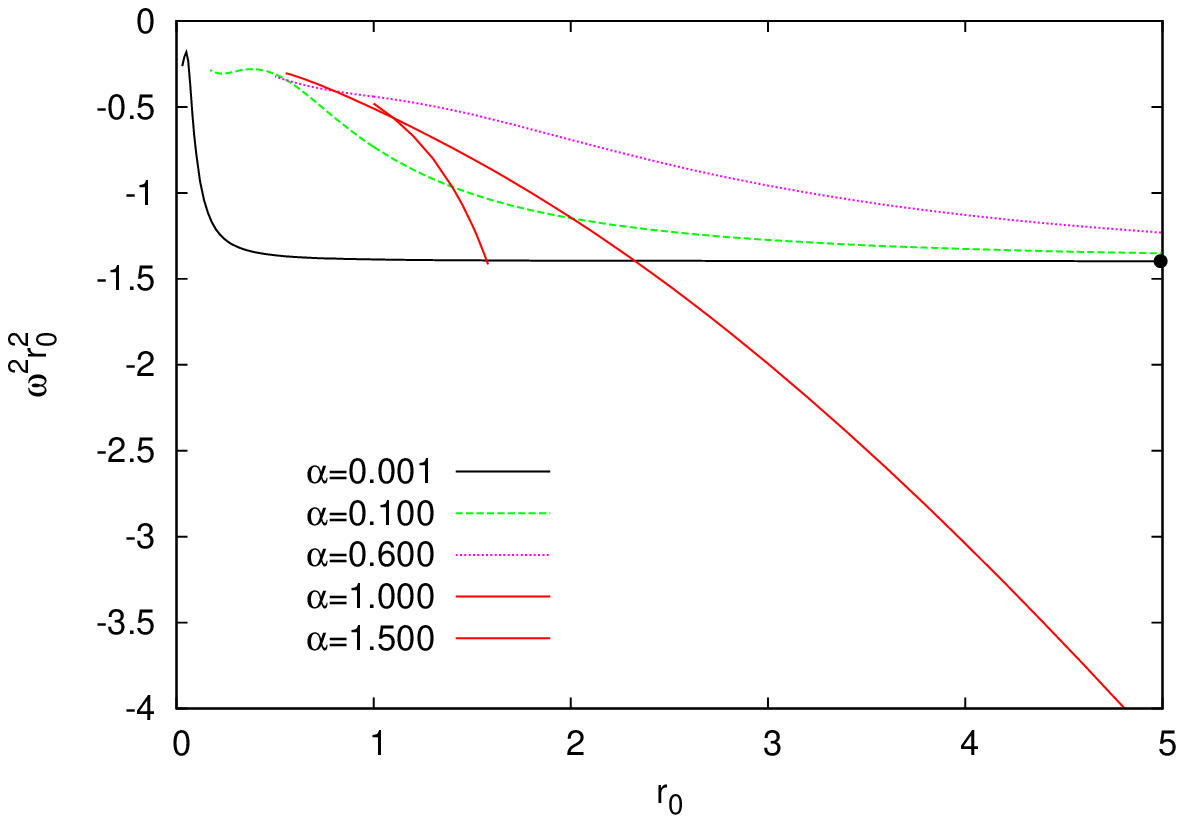}
\label{fig8b}
}
}
\end{center}\vspace{-0.7cm}
\caption{
Plots of (a) the scaled eigenvalue for the NLS wormholes as a function of $\alpha$ and
(b) 
the scaled eigenvalue for the Skyrmionic wormholes  as a function of $r_0$ for  different 
values of $\alpha$. 
The dot in (a) and (b) indicates the eigenvalue of the Ellis wormhole.
\label{fig8}
}
\end{figure}

The numerical results are demonstrated in Fig. \ref{fig8}. 
The case of the NLS wormholes is displayed in Fig. \ref{fig8a}, 
where the eigenvalue is shown as a function of $\alpha$. Here the dot indicates
the eigenvalue of  the Ellis wormhole \cite{Gonzalez:2008wd}.
We observe that $\omega^2$ is 
negative and decreases with increasing $\alpha$. Thus we conclude that all  
NLS wormholes  are unstable.
In Fig. \ref{fig8b} the stability analysis of the Skyrmionic wormholes is summarized.
In particular,  the eigenvalue as a function of $r_0$ for several values of $\alpha$ is shown.
As in the case of NLS  wormholes, we conclude that $\omega^2$ is negative for 
all solutions. Hence  the Skyrmionic wormholes are unstable as well.

\section{Non-symmetric Skyrmionic Wormhole Solutions}

\begin{figure}[t!]
\begin{center}
\vspace{0.5cm}
\mbox{\hspace{-0.5cm}
\subfigure[][]{\hspace{-1.0cm}
\includegraphics[height=.25\textheight, angle =0]{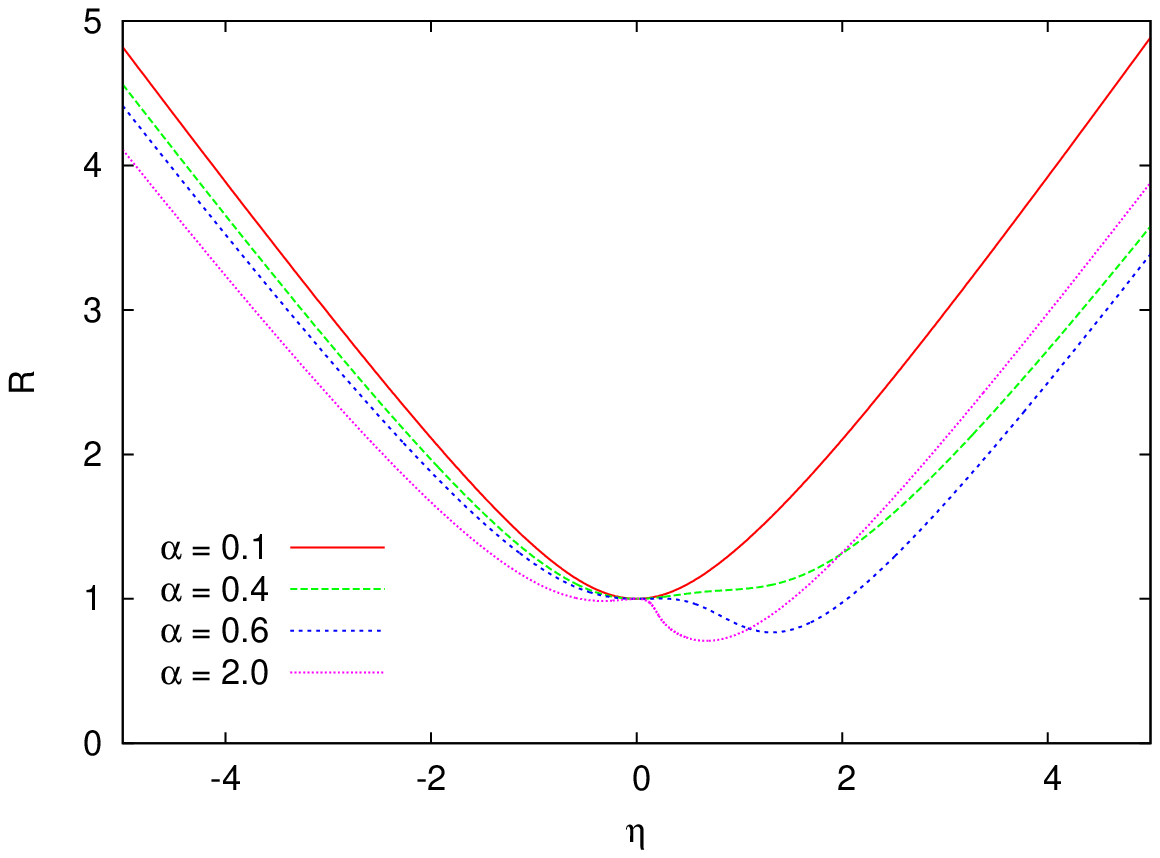}
\label{fig9a}
}
\subfigure[][]{\hspace{-0.5cm}
\includegraphics[height=.25\textheight, angle =0]{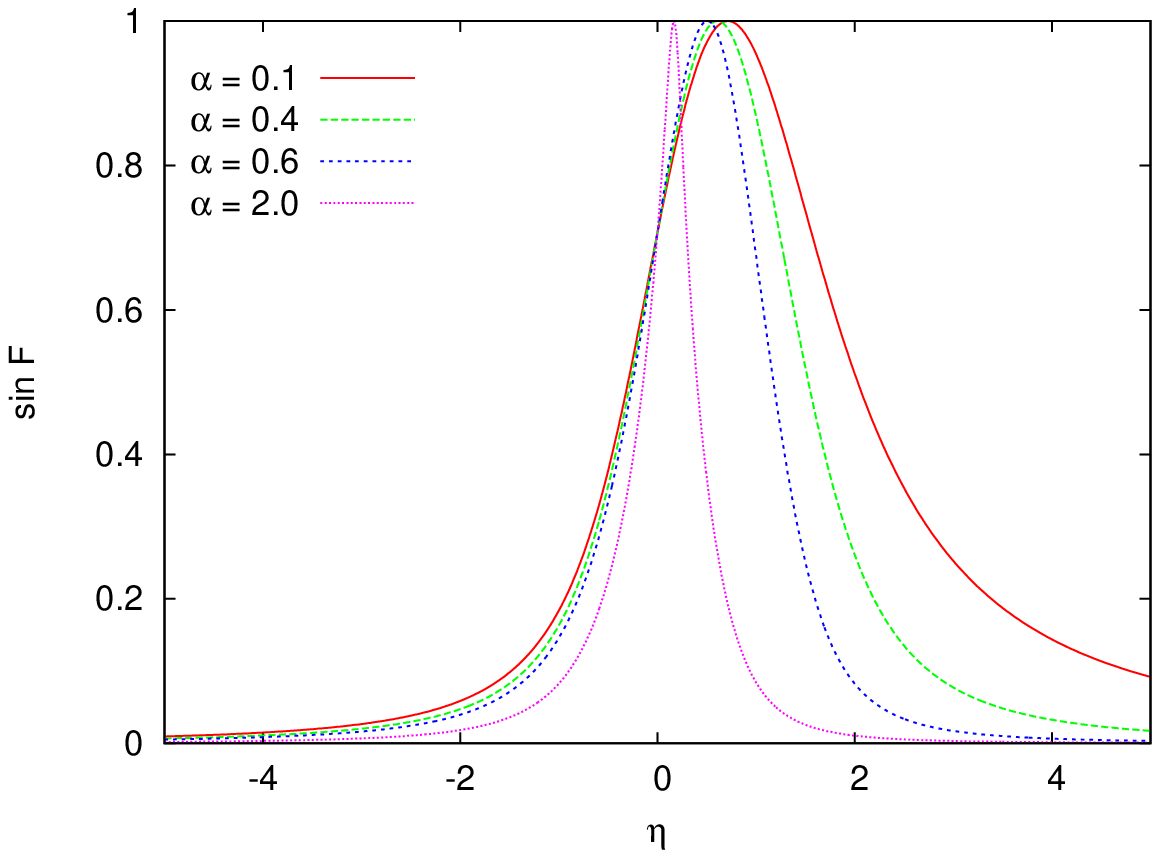}
\label{fig9b}
}
}
\mbox{\hspace{-0.5cm}
\subfigure[][]{\hspace{-1.0cm}
\includegraphics[height=.25\textheight, angle =0]{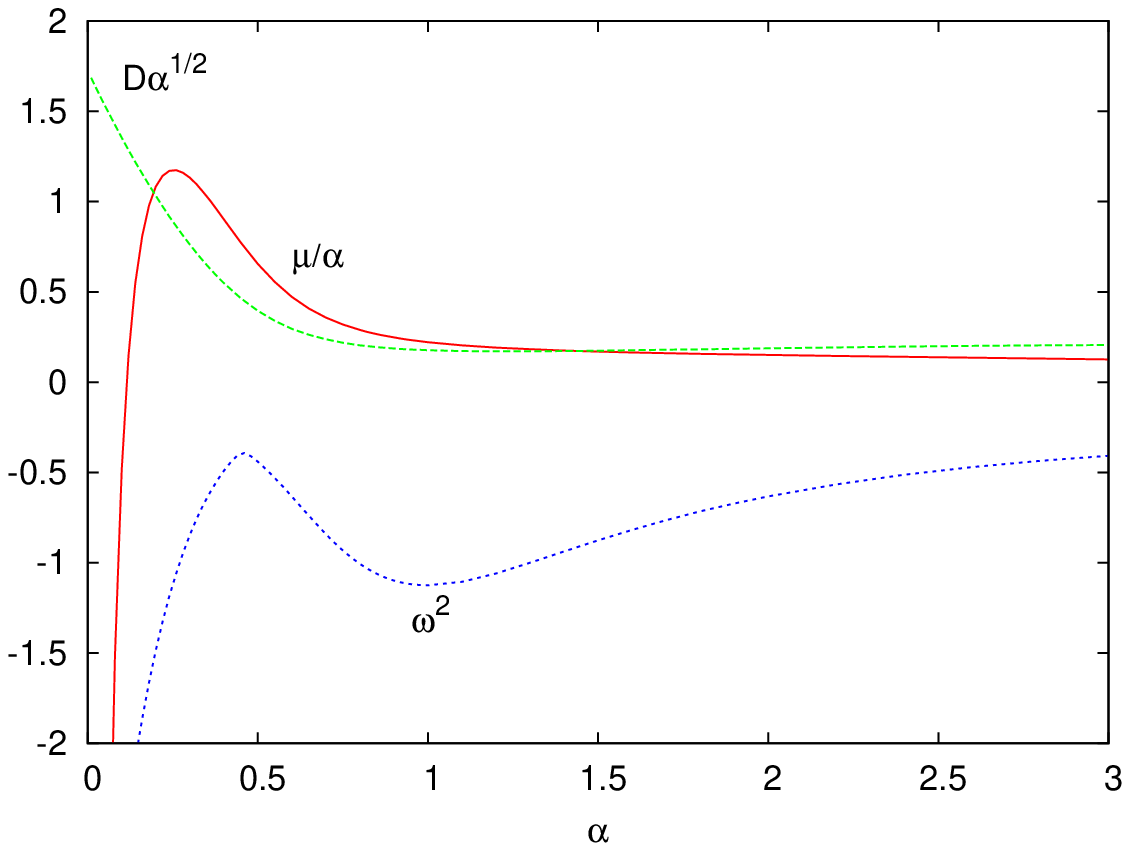}
\label{fig9c}
}
}
\end{center}\vspace{-0.7cm}
\caption{
Plots of (a) the areal radius $R(\eta)$,
(b) the function $\sin F(\eta)$,
and (c) the dimensional mass $\mu/\alpha$ and scalar charge $D/\sqrt{\alpha}$
as well as the unstable mode for the non-symmetric Skrymionic wormholes
for different values of $\alpha$ when $r_0=1$ and $F_0 = 3\pi/4$.
\label{fig9}
}
\end{figure}

The wormholes we considered so far possess the symmetry
$A(-\eta)=A(\eta)$, $R(-\eta)=R(\eta)$ and $F(-\eta)= \pi-F(\eta)$.
The last property implies $F(0)=\pi/2$. However, since $F(0)=F_0$ is a free parameter
we can choose $F_0\neq \pi/2$. Such a choice breaks the symmetry 
with respect to $\eta \to -\eta$  
and leads  to {\it non-symmetric} wormholes.
As an example we consider the non-symmetric Skyrmionic wormholes 
for fixed parameters $F_0= 3\pi/4$ and $r_0=1$ and varying coupling parameter 
$\alpha$. The areal radius $R$ and the function $\sin F$ are shown for several
values of $\alpha$ in Figs.~\ref{fig9a} and \ref{fig9b}, respectively.
As can be seen from Fig.~\ref{fig9a}, the asymmetry of the areal function $R$ is larger for large $\alpha$.
In contrast, Fig. ~\ref{fig9b} shows that the asymmetry of the chiral field
is more pronounced for small values of $\alpha$.   
However, one should keep in mind that $\alpha$ is 
the coupling strength in the Einstein equations.
Thus, for small $\alpha$ the asymmetry of the chiral field 
does not affect the space-time metric very much, leading to an almost
symmetric function $R$. On the other hand,  
for large $\alpha$ the asymmetry of the chiral field has considerable effect
on the metric and induces an asymmetry in the metric functions. From Fig.~\ref{fig9b} we 
observe that with increasing $\alpha$ the minimum
of the areal function turns to a maximum and two local minima next to it.
This is similar to the symmetric solutions, however the depths of the two minima
differ and they are not located symmetrically around the maximum.
These non-symmetric wormhole solutions thus also possess two non-symmetric throats.

Fig.~\ref{fig9c} displays the scaled mass parameter $\mu/\alpha$ and the 
scaled scalar charge $D\sqrt{\alpha}$ in terms of $\alpha$.
We observe that $\mu/\alpha$  assumes negative values for small $\alpha$,  
possesses a maximum at $\alpha \approx 0.26$, and tends to some finite value
for large $\alpha$.
In addition,  Fig.\ref{fig9c} displays the eigenvalue $\omega^2$ of the unstable 
mode. Since  $\omega^2$ is negative we conclude that the non-symmetric
Skyrmionic wormholes (as considered so far) are, also, unstable.
  
\section{Conclusion}

We have considered Morris-Thorne wormholes threaded by chiral fields
that carry a conserved topological charge.
We have focussed on static spherically symmetric solutions,
which are symmetric under an interchange of the
two asymptotically flat universes with respect to the throat.

When the chiral fields are described by the NLS model,
the coupling parameter $\alpha$  is the single free parameter for symmetric
wormholes. For small $\alpha$ the chiral fields have hardly any influence
upon the wormhole geometry, and the solutions possess a single
throat.
As the coupling parameter increases, the presence of the chiral fields
starts to be felt by the geometry. 
At a critical coupling $\alpha_{\rm cr}$, the throat becomes degenerate.
Beyond $\alpha_{\rm cr}$ the wormhole then exhibits a set of three extrema,
a (local) maximal surface  and two minimal surfaces located symmetrically,
one on each side. Thus the wormhole possesses a double throat
with a belly in the interior.

As $\alpha$ increases further 
the chiral fields become more and more localized in the inner region of the belly.
At the maximal coupling $\alpha_{\rm max}$
finally, the chiral fields are fully localized inside this inner region,
while the areal radius of the two throats decreases to zero.
The space-time then splits into three parts.
The outer parts correspond to empty Minkowski spaces, while the inner part
corresponds to an Einstein universe with a chiral field
carrying topological charge one.

In the presence of a Skyrme term, we have one more free parameter.
Thus we can vary the gravitational coupling and the throat size independently.
With increasing $\alpha$ we again observe an increasing influence 
of the matter on the geometry of the throat,
and the formation of two throats with
an inner belly beyond a critical value $\alpha_{\rm cr}$.
The splitting of the space-time into three parts is, however, less smooth
than in the NLS case.

By choosing the boundary conditions asymmetrically with respect to the throat
we can distribute the chiral fields asymmetrically with respect to the two asymptotically
flat universes. 
Then the deformation of the wormhole by the matter is asymmetric as well.  
Consequently, when 
the single throat splits into a double throat at a critical coupling,
the two throats develop asymmetrically with respect to their location and size.
This may possibly lead to only a
single throat reaching zero, resulting in
a breakup of the space-time into only two parts at a maximal coupling.

Let us now think of the Skyrmionic wormholes as 
Skyrmions passing a wormhole. Clearly, one would have to make
time-dependent calculations to model such a transit.
However, the static calculations may already give us some hints
of what to expect.
In particular, we conclude that the Skyrmions will 
deform the wormhole geometry
and possibly even let the space-time pinch,
resulting in two or three disconnected parts.

Let us conclude by stating that the chiral fields cannot stabilize the wormhole space-time.
The instability of the Ellis solution is inherited by the wormholes threaded by
chiral fields, although they {\it do}  carry a topological charge.
According to our observations this is not surprising though,
since the topological charge may finally simply reside
in a single of several disconnected space-time parts.

\section*{Acknowledgement}

B. Kleihaus and J. Kunz gratefully acknowledge support by the German Research Foundation  
within the framework of the DFG Research Training Group 1620 {\it Models of gravity}.
E. Charalampidis acknowledges financial support by the AUTH Research Committee.


\begin{thebibliography}{99}
  

\bibitem{Visser:1995cc}
For an overview see e.~g.
  M.~Visser,
  ``Lorentzian wormholes: From Einstein to Hawking'',
  Woodbury, USA: AIP (1995) 412 p
  
\bibitem{Einstein:1935tc}
  A.~Einstein and N.~Rosen,
  Phys.\ Rev.\  {\bf 48} (1935) 73.


\bibitem{Wheeler:1957mu}
  J.~A.~Wheeler,
  Annals Phys.\  {\bf 2}, 604-614 (1957).

\bibitem{Wheeler:1962} 
  J.~A.~Wheeler, 
  {\it Geometrodynamics} (Academic, New York, 1962).

\bibitem{Kruskal:1959vx}
  M.~D.~Kruskal,
  Phys.\ Rev.\  {\bf 119}, 1743-1745 (1960).
  
\bibitem{Fuller:1962zza}
  R.~W.~Fuller, J.~A.~Wheeler,
  Phys.\ Rev.\  {\bf 128}, 919-929 (1962).
  

\bibitem{Redmount:1985}
  I.~H.~Redmount, 
  Prog.\ Theor.\ Phys.\ {\bf 73} (1985) 140.

\bibitem{Eardley:1974zz}
  D.~M.~Eardley,
  Phys.\ Rev.\ Lett.\  {\bf 33}, 442-444 (1974).

\bibitem{Wald:1980nk}
  R.~M.~Wald, S.~Ramaswamy,
  Phys.\ Rev.\  {\bf D21}, 2736-2741 (1980).
  
\bibitem{Morris:1988cz}
  M.~S.~Morris, K.~S.~Thorne,
  Am.\ J.\ Phys.\  {\bf 56}, 395-412 (1988).
  
\bibitem{Ellis:1973yv}
  H.~G.~Ellis,
  J.\ Math.\ Phys.\  {\bf 14}, 104-118 (1973).

\bibitem{Ellis:1979bh}
  H.~G.~Ellis,
  Gen.\ Rel.\ Grav.\  {\bf 10}, 105-123 (1979).

\bibitem{Bronnikov:1973fh}
  K.~A.~Bronnikov,
  Acta Phys.\ Polon.\  {\bf B4}, 251-266 (1973).

\bibitem{Kodama:1978dw}
  T.~Kodama,
  Phys.\ Rev.\  {\bf D18}, 3529-3534 (1978).

\bibitem{ArmendarizPicon:2002km}
  C.~Armendariz-Picon,
  Phys.\ Rev.\  {\bf D65}, 104010 (2002)
  [gr-qc/0201027].

\bibitem{Dzhunushaliev:2011xx}
  V.~Dzhunushaliev, V.~Folomeev, B.~Kleihaus and J.~Kunz,
  JCAP {\bf 1104}, 031 (2011)
  [arXiv:1102.4454 [astro-ph.GA]].

\bibitem{Dzhunushaliev:2012ke}
  V.~Dzhunushaliev, V.~Folomeev, B.~Kleihaus and J.~Kunz,
  Phys.\ Rev.\ D {\bf 85}, 124028 (2012)
  [arXiv:1203.3615 [gr-qc]].

\bibitem{newpap}
  V.~Dzhunushaliev, V.~Folomeev, B.~Kleihaus and J.~Kunz,
  ``Mixed neutron-star-plus-wormhole systems: Linear stability analysis'',
  [arXiv:1302.5217 [gr-qc]].

\bibitem{Hochberg:1990is}
  D.~Hochberg,
  Phys.\ Lett.\  {\bf B251}, 349-354 (1990).

\bibitem{Fukutaka:1989zb}
  H.~Fukutaka, K.~Tanaka, K.~Ghoroku,
  Phys.\ Lett.\  {\bf B222}, 191-194 (1989).

\bibitem{Ghoroku:1992tz}
  K.~Ghoroku, T.~Soma,
  Phys.\ Rev.\  {\bf D46}, 1507-1516 (1992).

\bibitem{Furey:2004rq}
  N.~Furey, A.~DeBenedictis,
  Class.\ Quant.\ Grav.\  {\bf 22}, 313-322 (2005).
  [gr-qc/0410088].

\bibitem{Bronnikov:2009az}
  K.~A.~Bronnikov and E.~Elizalde,
  Phys.\ Rev.\  D {\bf 81}, 044032 (2010)
  [arXiv:0910.3929 [hep-th]].


\bibitem{Kanti:2011jz} 
  P.~Kanti, B.~Kleihaus and J.~Kunz,
  Phys.\ Rev.\ Lett.\  {\bf 107}, 271101 (2011)
  [arXiv:1108.3003 [gr-qc]].


\bibitem{Kanti:2011yv}
  P.~Kanti, B.~Kleihaus and J.~Kunz,
  Phys.\ Rev.\ D {\bf 85} (2012) 044007
%
  [arXiv:1111.4049 [hep-th]].
  
\bibitem{Luckock:1986tr} 
  H.~Luckock and I.~Moss,
 Phys. Lett. {\bf B176} (1986) 341;\\
 H. Luckock,
 {\it Black hole skyrmions},
 Proceedings of the 1986 Paris-Meudon Colloquium,
 eds. H.~J. de Vega, and N. Sanchez,
 (World Scientific, Singapore, 1987).

\bibitem{Droz:1991cx} 
  S.~Droz, M.~Heusler and N.~Straumann,
  Phys.\ Lett.\ B {\bf 268}, 371 (1991).


\bibitem{Gonzalez:2008wd}
  J.~A.~Gonzalez, F.~S.~Guzman and O.~Sarbach,
  Class.\ Quant.\ Grav.\  {\bf 26} (2009) 015010
  [arXiv:0806.0608 [gr-qc]].
  
\bibitem{Bronnikov:2011if}
  K.~A.~Bronnikov, J.~C.~Fabris and A.~Zhidenko,
  Eur.\ Phys.\ J.\ C {\bf 71} (2011) 1791
  [arXiv:1109.6576 [gr-qc]].

\bibitem{Heusler:1991xx} 
  M.~Heusler, S.~Droz and N.~Straumann,
  Phys.\ Lett.\ B {\bf 271}, 61 (1991).

\bibitem{Heusler:1992av} 
  M.~Heusler, S.~Droz and N.~Straumann,
  Phys.\ Lett.\ B {\bf 285}, 21 (1992).


\bibitem{Lechner:2000bw}
This solution has been discussed before in
  C.~Lechner, S.~Husa and P.~C.~Aichelburg,
  Phys.\ Rev.\ D {\bf 62} (2000) 044047
  [gr-qc/0002031].

\end{thebibliography}
\end{document}